\documentclass[authoryear,final,5p,times,twocolumn]{elsarticle}
\usepackage{amsmath}
\usepackage{chngpage}
\usepackage{multirow}
\usepackage{gensymb}
\usepackage{array,arydshln}
\usepackage{graphicx}
\usepackage[T1]{fontenc}
\usepackage[utf8]{inputenc}
\usepackage{setspace}
\usepackage{lineno}
\usepackage{hyperref}
\usepackage{amssymb}
\usepackage[font=normal]{caption}

\begin{document}

\begin{frontmatter}
  
\title{Obliquity and precession as pacemakers of Pleistocene deglaciations}
\author{Fabo Feng\corref{cor1}}
\cortext[cor1]{Corresponding author}
\ead{fengfabo@gmail.com}
\ead[url]{http://www.mpia.de/homes/ffeng/}
\author{C. A. L.\ Bailer-Jones\corref{cor2}}
\cortext[cor2]{Principal corresponding author}
\ead{calj@mpia.de}
\ead[url]{http://www.mpia.de/~calj/}
\address{{\it\small Max Planck Institute for Astronomy, K\"{o}nigstuhl 17, 69117 Heidelberg, Germany}\\
 \vspace{0.08in}
{\rm\small to appear in Quaternary Science Reviews (submitted 23 December 2014; accepted 7 May 2015)}}

\begin{abstract}
The Milankovitch theory states that the orbital eccentricity, precession, and obliquity of the Earth influence our climate by modulating the summer insolation at high latitudes in the northern hemisphere. Despite considerable success of this theory in explaining climate change over the Pleistocene epoch (2.6 to 0.01 Myr ago), it is inconclusive with regard to which combination of orbital elements paced the 100\,kyr glacial-interglacial cycles over the late Pleistocene. 
Here we explore the role of the orbital elements in pacing the Pleistocene deglaciations by modeling ice-volume variations in a Bayesian approach. When comparing models, this approach takes into account the uncertainties in the data as well as the different degrees of model complexity. We find that the Earth's obliquity (axial tilt) plays a dominant role in pacing the glacial cycles over the whole Pleistocene, while precession only becomes important in pacing major deglaciations after the transition of the dominant period from 41\,kyr to 100\,kyr (the mid-Pleistocene transition). We also find that geomagnetic field and orbital inclination variations are unlikely to have paced the Pleistocene deglaciations. We estimate that the mid-Pleistocene transition took place over a 220\,kyr interval centered on a time 715\,kyr ago, although the data permit a range of 600--1000\,kyr. This transition, occurring within just two 100\,kyr cycles, indicates a relatively rapid change in the climate response to insolation.
\end{abstract}    

\begin{keyword}
glacial cycles; obliquity; Pleistocene; Bayesian inference; mid-Pleistocene transition; climate model 
\end{keyword}
\end{frontmatter}

\section{Introduction}\label{sec:introduction}

During the past 1\,Myr (the late Pleistocene), the polar ice sheets grew slowly (glaciation) then retreated abruptly (deglaciation or glacial termination) repeatedly, with an interval of about 100\,kyr \citep{hays76}. These quasi-periodic glacial-interglacial cycles dominated terrestrial climate change. They are recorded by paleoclimatic proxies such as $\delta^{18}$O (the scaled $^{18}{\rm O}/^{16}{\rm O}$ isotope ratio) in foraminiferal calcite, which is sensitive to changes in global ice volume and ocean temperature. Following on from the work of Adh\'emar, Croll, and others, Milankovitch proposed that climate change is driven by the insolation (the received solar radiation) during the northern hemisphere summer at northerly latitudes \citep{milankovitch41}.  This insolation depends on the Earth's orbit and axial tilt (obliquity), and Milankovitch suggested that through various climate response mechanisms, variations in these orbital elements -- in particular eccentricity, obliquity, and precession\footnote{This involves both the orbital and the axial precession.} -- can cause climate change (``Milankovitch forcing'').  Many studies have broadly confirmed Milankovitch's theory and the role of Milankovitch forcing in driving Pleistocene climate change, for example by spectral analyses of paleoclimatic time series derived from deep-sea sediments \citep{hays76, shackleton73,kominz79}. These studies have demonstrated that the climate variance is concentrated in periods of about 19\,kyr, 23\,kyr, 42\,kyr and 100\,kyr which are close to the dominant periods in precession ($\sim$23 and 19\,kyr), obliquity ($\sim$41\,kyr), and eccentricity ($\sim$100 and 400\,kyr).

\begin{figure*}[ht!]
  \centering
  \includegraphics[scale=0.8]{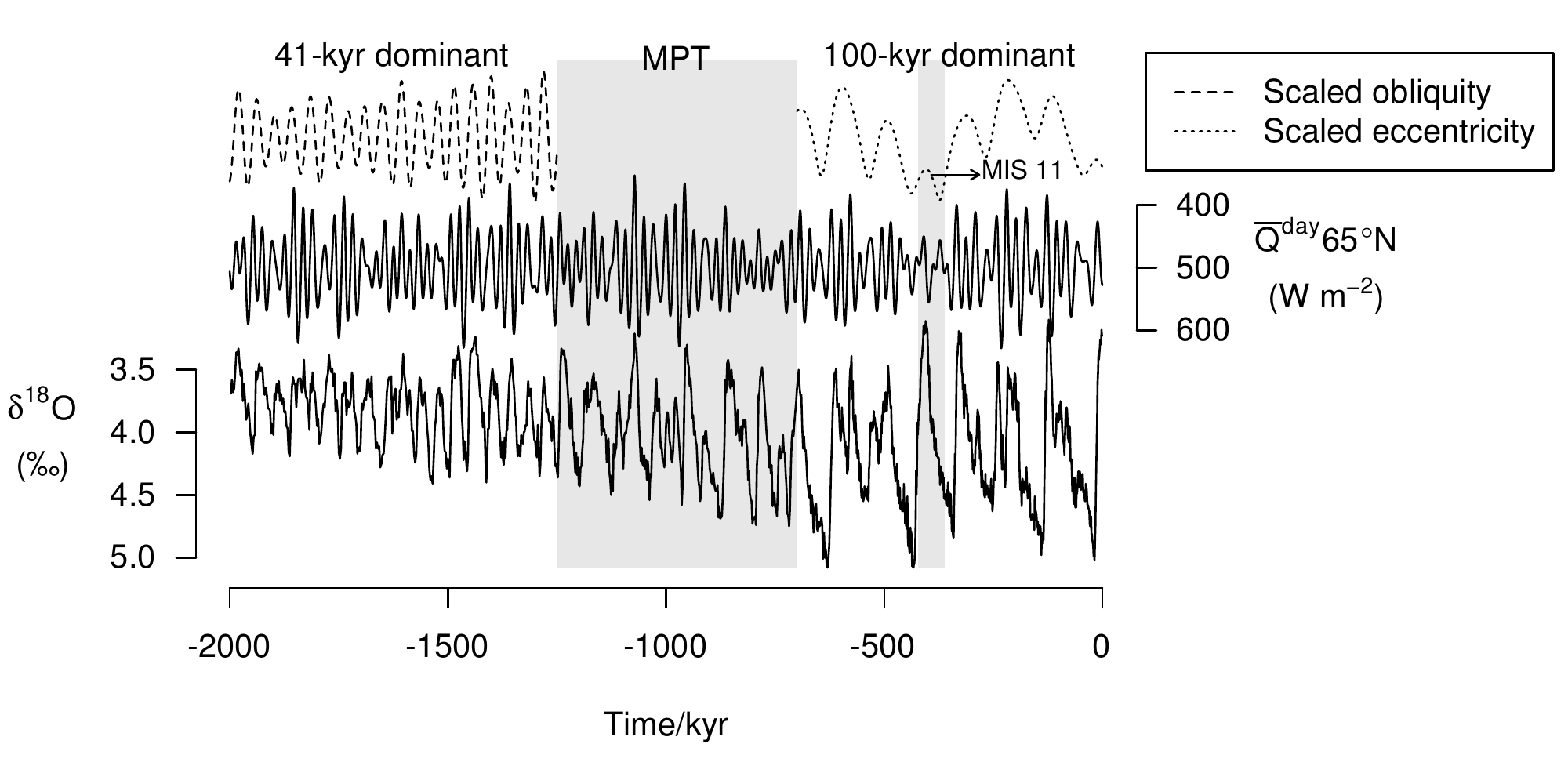}
  \caption{Climate variations over the Pleistocene. The present day is at time zero on the right. The $\delta^{18}$O record (lower solid line) stacked by \cite{lisiecki05} is compared with the daily-averaged insolation at the summer solstice at $65^{\circ}N$, $\bar{Q}^{\rm day}65^{\circ}N$ (upper solid line), the  obliquity (dashed line), and the eccentricity (dotted line) calculated by \cite{laskar04}. The latter two have been scaled to have a common amplitude. The grey region around $-1000$\,kyr represents the MPT extending from $-1250$\,kyr to $-770$\,kyr \citep{clark06}. The grey bar extending from $-423$ to $-362$\,kyr represents marine isotope stage (MIS) 11. The $\delta^{18}$O variations are dominated by 41\,kyr and 100\,kyr cycles before and after the MPT respectively. }
  \label{fig:milankovitch_diagram}
\end{figure*}

There are, however, several difficulties in reconciling the Milankovitch theory with observation. Two in particular arise when trying to explain the 100\,kyr cycles. The first is the transition from the 41\,kyr dominant period in climate variations to a 100\,kyr dominant period at the mid-Pleistocene around 1\,Myr ago (hereafter ``Myr ago'' is written ``Ma''). The second difficulty is generating 100\,kyr sawtooth variations from orbital forcings and climate response mechanisms (\citealt{imbrie93}, \citealt{huybers07}, \citealt{lisiecki10}). On the one hand, and as shown in Figure \ref{fig:milankovitch_diagram}, the onset of 100\,kyr power at the mid-Pleistocene transition (MPT) occurs without a corresponding change in the summer insolation at high northern latitudes (represented by the daily-averaged insolation on 21 June at $65^{\circ}$N). On the other hand, the $\sim$100\,kyr eccentricity cycle produces only negligible 100\,kyr power in seasonal or mean annual insolation variations, despite its modulation of the precession amplitude. Furthermore, the variations of eccentricity and the northern summer insolation are weak while the 100\,kyr climatic variations are strong, notably in marine isotope stage (MIS) 11 (see Figure \ref{fig:milankovitch_diagram} and \citet{imbrie80,howard97}). These problems are referred to as the ``100\,kyr problem'' \citep{imbrie93}.

Various models with different climate forcings and response mechanisms have been proposed to solve the 100\,kyr problem. Many are based on either deterministic climate forcing models or stochastic internal climate variations. The former proposes that the 100\,kyr cycles are driven by orbital variations, particularly precession and eccentricity \citep{imbrie80,paillard98,gildor00}. Many models treat the insolation variation as a pacemaker which sets the phase of the glacial-interglacial oscillation by directly controlling summer melting of ice sheets \citep{gildor00}. In this latter hypothesis, stochastic internal climate variability plays the main role in generating the 100\,kyr glacial cycles \citep{saltzman82,pelletier03,wunsch03}. A general approach is to combine the deterministic and stochastic elements within a framework of nonlinear dynamics, which allows for the occurrence of bifurcation and synchronisation in the climate system (see review by \citealt{crucifix12b}).

Other proposed hypotheses include glaciation cycles controlled by the accretion of interplanetary dust when the Earth crosses the invariable plane \citep{muller97} or by the cosmic ray flux modulated by the Earth's magnetic field (measured as the geomagnetic paleointensity, GPI; \citealt{christl04,courtillot07}). Some models also try to explain the MPT with \citep{raymo97, paillard98,honisch09,clark06} or without \citep{huybers09,lisiecki10,imbrie11} an internal change in the climate system.

The above models comprise both climate forcings and responses. According to various studies \citep{saltzman87,maasch90,ghil94,raymo97,paillard98,clark99,tziperman03,ashkenazy04}, climate forcings frequently determine the time of occurrence of some climate feature, such as the onset of deglaciation.  
Many recent studies have employed concepts from chaos theory to address the problem of climate change \citep{crucifix12b,parrenin12,crucifix13,mitsui14,ashwin15,williamson15}, which then allow the concept of "pacing" to be described more rigorously as a forcing mechanism.  \cite{huybers11} noted that many tens of pacing models have been proposed, yet we lack the means to choose between them.

Our current work aims to compare different forcing mechanisms by using a simple ice volume model for the Pleistocene glacial-interglacial cycles. We adopt the pacing model given by \cite{huybers05} and combine it with different forcings in order to predict the glacial terminations, which are identified from several $\delta^{18}$O records. Our models do not describe the physical mechanism of the climate response to external forcings. We aim instead only to measure the role of different forcings in determining the times of deglaciations. Due to the large and rapid change in ice volume at deglaciation, these times are relatively easy to identify, so the time uncertainties associated with identification are small. They are nonetheless still affected by the overall uncertainty in the chronology of the $\delta^{18}$O record \citep{huybers05}.

A common approach for assessing a model is to use p-values to reject a null hypotheses \citep{huybers05,huybers11}. However, it is well established that p-values can give very misleading results \citep{berger87,jaynes03,christensen05,bailer-jones09,feng13}, so we instead compare models using the Bayesian evidence. This compares models on an equal footing and takes into account the different flexibility (or complexity) of the models \citep{kass95,spiegelhalter02,vonToussaint11}.

This paper is organized as follows. In section \ref{sec:data} we assemble the data -- stacked $\delta^{18}$O records -- and identify the glacial terminations. In section \ref{sec:bayes} we summarize the Bayesian inference method as we use it. We build models based primarily on orbital elements to predict the Pleistocene glacial terminations in section \ref{sec:model}. These are compared for different data sets and time scales in section \ref{sec:comparison}.  We perform a test of sensitivity of the results to the model parameters and choice of time scales in section \ref{sec:sensitivity}. Finally, we discuss our results and conclude in section \ref{sec:conclusion}.

\section{Data}\label{sec:data}

\subsection{$\delta^{18}$O from a depth-derived age model}\label{sec:delta18O}

The past climate can be reconstructed from isotopes recorded in ice cores or deep sea sediment cores. Air bubbles trapped at different depths in ice cores can be used to reconstruct the past atmospheric temperature, for example. Ice cores have so far been used to trace the climate back to about 800\,kyr \citep{augustin04}.  In order to reconstruct the climate back to 2\,Ma, the $\delta^{18}$O ratio recorded in the calcite (CaCO$_3$) in foraminifera fossils (including species of benthos and plankton) in ocean sediment cores can be used.  We use the $\delta^{18}$O ratio as a measure of variations in the global ice volume, although we note that this is also sensitive to the temperature and isotope composition of seawater, for which corrections can be made. For a discussion of the interpretation of marine calcite $\delta^{18}$O see for example \cite{shackleton67} and \cite{mix84}.

In order to calibrate $\delta^{18}$O measurements and to assign ages to sediment cores, one could assume either a constant sedimentation rate (determined using radiometrically dated geomagnetic reversals), or a constant phase relationship between $\delta^{18}$O and an insolation forcing based on the Milankovitch theory (see \citealt{huybers04} for details). The former is the ``depth-derived age model'' \citep{huybers04, huybers07}. The latter is referred to as ``orbital tuning'' \citep{imbrie84,martinson87,shackleton90}. Clearly this latter method is not appropriate for testing theories related to Milankovitch forcings, because it already assumes a link between $\delta^{18}$O variations and orbital forcings. 

\cite{huybers07} (hereafter H07) stacked and averaged twelve benthic and five planktic $\delta^{18}$O records to generate three $\delta^{18}$O global records: an average of all $\delta^{18}$O records (``HA'' data set); an average of the benthic records (``HB'' data set); an average of the planktic records (``HP'' data set).\footnote{The planktic $\delta^{18}$O records may not produce a stack as good as benthic records because surface water is less uniform in temperature and salinity than the deep ocean \citep{lisiecki05}.} In addition to these three data sets, we also analyze the orbital-tuned benthic $\delta^{18}O$ stacked by \cite{lisiecki05} (``LR04'' data set), despite its orbital assumptions. The LR04 record was re-calibrated by H07 to generate a tuning-independent LR04 data set (``LRH'' data set; see the supplementary material of H07 for details).

\begin{figure*}[ht!]
  \centering
  \includegraphics[scale=0.7]{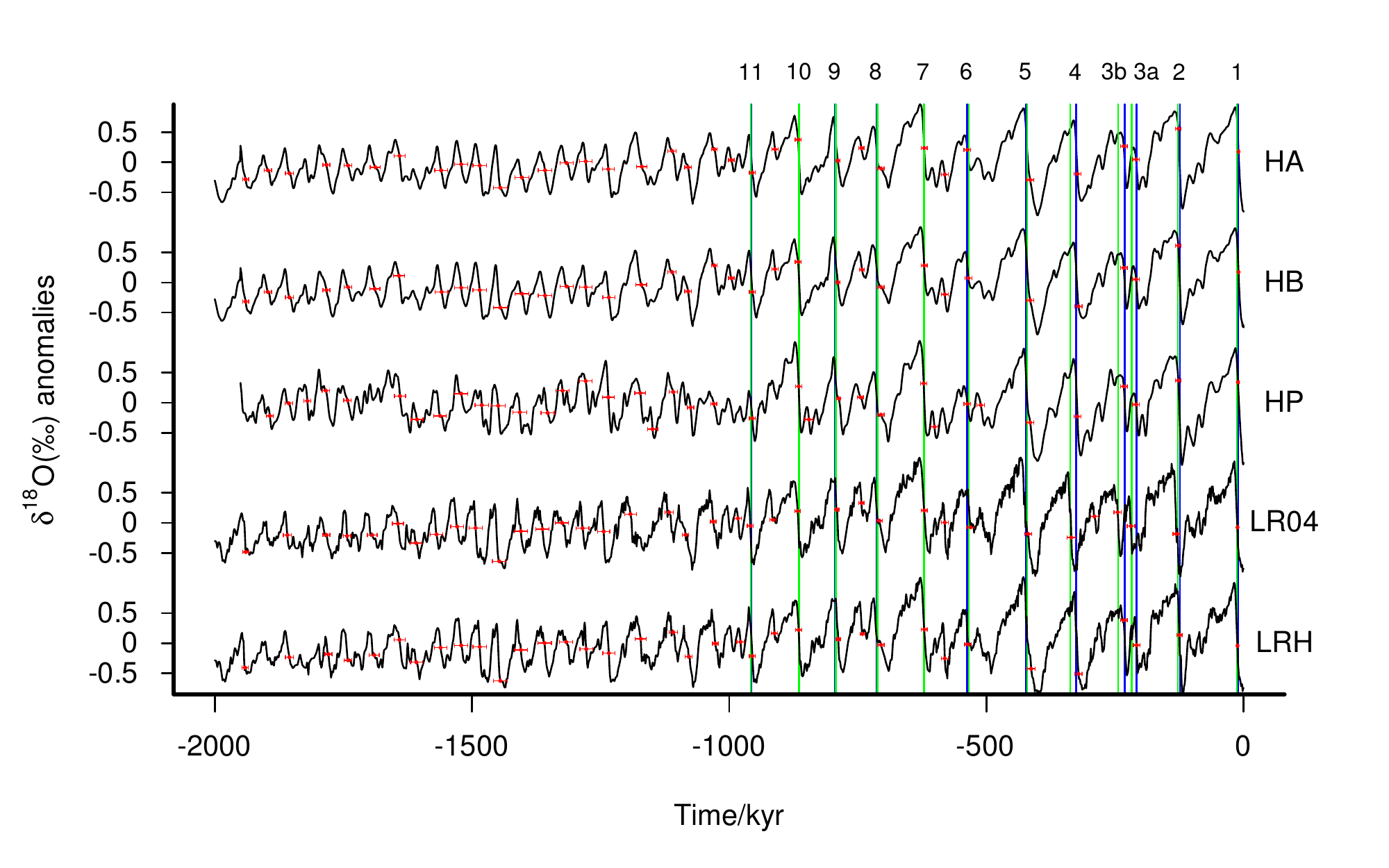}
  \caption{The variation of $\delta^{18}$O with time as determined by a depth-derived age-model (HA, HB, HP, and LRH) and an orbital-tuning model (LR04). The past 2000\,kyr is divided into two parts: the early Pleistocene extending from 2\,Ma to 1\,Ma and the late Pleistocene extending from 1\,Ma to the present. 
The deglaciations we identify for each data set are show in red: the point is the mean time, the error bar is the uncertainty. In the late Pleistocene, we identify three additional sets of terminations: the DD terminations are denoted by blue lines while the ML/MS terminations are denoted by green lines. These each consists of 12 major terminations, which are indicated by the numbers (we use the convention of splitting termination 3 into two events). What we call minor terminations are all the red points which are not major terminations.}
  \label{fig:huybers_data}
\end{figure*}

We standardize each of the above $\delta^{18}$O records over the past 2\,Myr to have zero mean and unit variance, to produce what we call the $\delta^{18}$O anomalies as shown in Figure \ref{fig:huybers_data} (DD, ML, MS are explained below).  We identify the deglaciations in the next section.  We see that the sawtooth 100\,kyr glacial-interglacial cycles become significant over the late Pleistocene while 41\,kyr cycles dominate over the early Pleistocene. From now on, we will use the term ``late Pleistocene'' to mean the time span 1\,Ma to 0\,Ma, and ``early Pleistocene'' to mean 2\,Ma to 1\,Ma.

\subsection{Identification of deglaciations}\label{sec:identification}

Rather than trying to model the full time series of $\delta^{18}$O variations, we focus instead only on the times of glacial terminations (deglaciations).
This is because an orbital forcing should determine predominantly the timing of a deglaciation rather than the detailed variation of the ice volume \citep{gildor00,paillard98,huybers05}. This not only simplifies the problem (thus making results more robust), but is also in line with our goal of trying to identify the main pacemakers for deglaciations, rather than trying to model the continuous response of the climate to orbital forcings. Here we describe how we identify the deglaciations.

From Figure \ref{fig:milankovitch_diagram}, we see that the $\delta^{18}$O amplitudes are larger in the late Pleistocene than in the early Pleistocene. This is interpreted to mean that after the MPT, ice sheets both grew to larger volumes and retreated more rapidly to ice-free conditions.
This rapid and abrupt shift from extreme glacial to extreme interglacial conditions defines 11 well-established late-Pleistocene terminations \citep{broecker84,raymo97b}. Because termination 3 is sometimes split into two terminations \citep{huybers11} -- labeled 3a and 3b (Figure \ref{fig:huybers_data}) -- we actually identify 12 {\it major terminations} over the late Pleistocene. The times of these major terminations as  established by various publications has been collated by \cite{huybers11} and are given in his supplementary material. Based on his Table S2, we define three sets of terminations which cover just the late Pleistocene:
\begin{itemize}
\item DD: termination times and corresponding uncertainties estimated from the depth-derived timescale in H07;
\item MS: termination times and corresponding uncertainty equal to the median and standard deviation (respectively) of different termination times for each event given in the literature \citep{imbrie84,shackleton90,lisiecki05,jouzel07,kawamura07};
\item ML: termination times as in the MS data set, but with larger uncertainties obtained by adding the time uncertainties of the depth-derived time scales in quadrature with the corresponding uncertainties in the MS data set.
\end{itemize}
These terminations are shown as vertical lines in Figure \ref{fig:huybers_data}. 

In addition to these major terminations, there are also minor terminations characterized by transitions from moderate glacial to moderate interglacial conditions. Considering the ambiguity in defining these \citep{huybers05,lisiecki10}, we identify terminations in our $\delta^{18}$O records using the method of H07. A termination is identified when a local maximum and the following minimum (defined as a maximum-minimum pair) have a difference in $\delta^{18}$O larger than one standard deviation of the whole $\delta^{18}$O record. The time of a termination is the mid-point of the maximum-minimum pair and the age uncertainty of this mid-point is calculated from a stochastic sediment accumulation rate model \citep{huybers07}. We identify sustained events in all data sets by filtering $\delta^{18}$O with different moving-average (or "Hamming") filters. The data sets are show in Figure \ref{fig:huybers_data}.  We use the term ``major terminations'' to refer to terminations identified in these data sets which coincide with the major terminations in the DD, MS, or ML data sets. All other terminations we refer to as minor terminations. The data on these are listed in Table \ref{tab:terminations}.

Finally, we also define three additional hybrid data sets. As the HA data set is a stack of both benthic and planktic records, we combine the early-Pleistocene terminations identified from the HA data set together with late-Pleistocene terminations from the DD, ML, and MS data sets to generate the HADD, HAML, and HAMS data sets, respectively.

Thus starting from our five original data sets (HA, HB, HP, LR04, LRH), we have a total of 11 data sets of glacial terminations against which we will compare our models (see Table \ref{tab:terminations}).

\begin{table*}
  \caption{Terminations (major and minor) identified from different $\delta^{18}$O records using H07's method (HA, HB, HP, LR04 and LRH) and the DD, MS and ML data sets of major terminations. Combining the early Pleistocene terminations of HA with the DD, MS and ML data sets, we obtain the hybrid data sets of HADD, HAMS and HAML. For each column, the termination ages are listed on the left side and the age uncertainties are listed on the right side (also see Figure \ref{fig:huybers_data}). All quantities are in units of kyr. }
  \label{tab:terminations}
  \centering
  \scalebox{0.9}{
    \begin{tabular}{|c|cc|cc|cc|cc|cc||cc|cc|cc|}
      \hline
    &\multicolumn{2}{|c|}{HA}&\multicolumn{2}{c|}{HB}&\multicolumn{2}{c|}{HP}&\multicolumn{2}{c|}{LR04}&\multicolumn{2}{c||}{LRH}&\multicolumn{2}{c|}{DD}&\multicolumn{2}{c|}{MS}&\multicolumn{2}{c|}{ML}\\\hline
    \multirow{17}{*}{{\parbox[t]{1.5cm}{Late\\Pleistocene\\(between 1\\and 0\,Ma)}}}&-10&0.81&-10&0.81&-11&1.9&-12&2.2&-12&2.2&-11&1.9&-13&1.8&-13&3.1\\
    &-127&5.3&-127&5.3&-127&5.3&-131&6.3&-125&5&-124&5&-128&3.6&-128&6.6\\
    &-209&6.6&-209&6.6&-209&6.6&-219&7.5&-208&6.4&-208&6.4&-218&4.3&-218&8.7\\
    &-233&6.4&-233&6.4&-233&6.4&-245&7&-233&6.4&-231&6.3&-244&4.8&-244&8.6\\
    &-323&6.8&-321&7&-323&6.8&-290&7.5&-321&7&-326&7&-337&4.5&-337&9.8\\
    &-415&7.4&-415&7.4&-415&7.4&-335&8.4&-413&7.6&-423&7.1&-421&4.4&-421&8.2\\
    &-537&6.5&-535&6.6&-537&6.5&-531&7.3&-581&6.9&-622&5.8&-621&2.7&-621&6.4\\
    &-581&6.9&-581&6.9&-537&6.5&-531&7.3&-581&6.9&-622&5.8&-621&2.7&-621&6.4\\
    &-621&5.8&-621&5.8&-601&6.4&-581&6.9&-621&5.8&-714&4.5&-712&7.5&-712&8.8\\
    &-705&5.9&-705&5.9&-622&5.8&-621&5.8&-705&5.9&-794&3.7&-793&1.8&-793&1.8\\
    &-743&5&-742&4.8&-705&5.9&-708&5.4&-741&4.5&-864&5.7&-864&0.84&-864&5.8\\
    &-789&4.2&-789&4.2&-745&5.5&-743&5&-788&4.2&-957&5.8&-958&1.7&-958&6.0\\
    &-866&5.8&-866&5.8&-787&4.1&-791&4.1&-865&5.7&&&&&&\\
    &-911&6&-911&6&-845&8&-867&5.7&-912&6&&&&&&\\
    &-955&5.9&-955&5.9&-865&5.7&-915&5.9&-955&5.9&&&&&&\\
    &-996&5.5&-996&5.5&-955&5.9&-959&5.7&-978&7&&&&&&\\
    &&   &     &   &     &   &-983&6.5  &   &&&&&&&\\
    \hline
    \multirow{21}{*}{{\parbox[t]{1.5cm}{Early\\Pleistocene\\(between 2\\and 1\,Ma)}}}&-1029&5.6&-1029&5.6&-1030&5.6&-1031&5.5&-1027&5.5&&&&&&\\
    &-1080&6.6&-1080&6.6&-1075&6.1&-1085&6.5&-1079&6.5&&&&&&\\
    &-1111&8.1&-1111&8.1&-1109&8&-1117&8&-1109&8&&&&&&\\
    &-1170&10.4&-1171&10.5&-1149&9.9&-1192&11.4&-1172&10.5&&&&&&\\
    &-1235&11.7&-1234&11.7&-1173&10.5&-1244&12&-1234&11.7&&&&&&\\
    &-1279&12.3&-1279&12.3&-1235&11.7&-1285&12.3&-1278&12.3&&&&&&\\
    &-1316&12.9&-1316&12.9&-1279&12.3&-1325&12.7&-1317&13&&&&&&\\
    &-1358&13.2&-1358&13.2&-1324&12.7&-1363&13.1&-1359&13.2&&&&&&\\
    &-1403&13.3&-1403&13.3&-1353&13&-1405&13.2&-1405&13.2&&&&&&\\
    &-1445&13.4&-1445&13.4&-1407&13.2&-1447&13.3&-1445&13.4&&&&&&\\
    &-1485&13.2&-1485&13.2&-1449&13.2&-1493&12.9&-1485&13.2&&&&&&\\
    &-1521&12.9&-1521&12.9&-1481&13.1&-1529&12.5&-1521&12.9&&&&&&\\
    &-1560&12.9&-1559&12.4&-1521&12.9&-1569&12&-1561&12.3&&&&&&\\
    &-1641&10.8&-1642&10.8&-1562&12.3&-1609&11.5&-1608&11.5&&&&&&\\
    &-1688&9.8&-1689&9.8&-1607&11.5&-1644&10.7&-1641&10.8&&&&&&\\
    &-1741&7.4&-1741&7.4&-1640&10.8&-1694&9.4&-1690&9.7&&&&&&\\
    &-1783&6.9&-1783&6.9&-1742&7.4&-1743&7.3&-1741&7.4&&&&&&\\
    &-1855&7.7&-1855&7.7&-1784&7&-1783&6.9&-1855&7.7&&&&&&\\
    &-1897&7.3&-1897&7.3&-1820&6.9&-1859&7.6&-1855&7.7&&&&&&\\
    &-1940&5.8&-1940&5.8&-1856&7.7&-1940&5.8&-1941&5.9&&&&&&\\
    &&&&&-1893&7.1&&&&&&&&&&\\
    \hline
  \end{tabular}}
\end{table*}

As there are dating errors and identification uncertainties, we cannot know exactly when a deglaciation occurred. To take into account these uncertainties, we treat the time of each deglaciation probabilistically by defining a Gaussian distribution with the mean and standard deviation equal to the time and time uncertainty (respectively) of the termination. The terminations in a data set are therefore represented as a sequence of Gaussians, which will be modeled as described in the following section. 

\section{Bayesian modelling approach}\label{sec:bayes}

We use the standard Bayesian probabilistic framework (e.g.\ \citealp{kass95,jeffreys61,mackay03,sivia2006}) to compare how well the different models explain the paleontological data. This approach takes into account the measurement errors, accounts consistently for the differing degrees of complexity present in our models, and compares models symmetrically.  Our specific methodology is outlined briefly in this section. It is described in more detail in \cite{bailer-jones11, bailer-jones11b}, where we also present arguments why this approach should be preferred to hypothesis testing using p-values.

The posterior probability of a model $M$ postulated to describe a data set $D$ is given by the rules of probability as
\begin{equation}
  P(M|D) = \frac{P(D|M)P(M)}{P(D)},  
\label{eqn:bayes1}
\end{equation}
where $P(M)$ is the prior of model $M$, and $P(D)$ can be considered here as a normalization constant. $P(D|M)$ is the {\it evidence} of model $M$ which can be written mathematically as
\begin{equation}
  P(D|M)=\int P(D|\boldsymbol{\theta},M)P(\boldsymbol{\theta}|M)d\boldsymbol{\theta} \ .
\label{eqn:bayes2} 
\end{equation}
$\boldsymbol{\theta}$ is the set of parameters of model $M$, $P(D|\boldsymbol{\theta},M)$ is the {\it likelihood} -- the probability of observing the data $D$ given specific values of the model parameters -- and $P(\boldsymbol{\theta}|M)$ is the {\em prior distribution} of parameters of this model.

Ideally we would be interested in evaluating the $P(M|D)$ for different models, as this is the probability of a model being true given the observed data. However, this would require that we define {\em all} possible models. Thus in practice we compare models by looking at the ratio of model posterior probabilities.
If we cannot (or choose not to) distinguish between models a priori, then we set $P(M)$ to be equal for all models. It follows from equations~\ref{eqn:bayes1} and \ref{eqn:bayes2} that this ratio for models $M_1$ and $M_2$ is
\begin{equation}
  \frac{P(M_1|D)}{P(M_2|D)} = \frac{P(D|M_1)}{P(D|M_2)}=\frac{\int P(D|\boldsymbol{\theta_1},M_1) P(\boldsymbol{\theta_1}|M_1)d \boldsymbol{\theta_1}}{\int P(D|\boldsymbol{\theta_2},M_2) P(\boldsymbol{\theta_2}|M_2)d \boldsymbol{\theta_2}}.
  \label{eqn:bayes3}
\end{equation}
The above ratio of the evidences is called the {\it Bayes factor} and is used to compare how well a model (relative to another model) predicts the data, independent of the values of the model parameters. Note that this does not involve tuning the model parameters, which is why using the evidence takes into account differing model complexities. A (maximum) likelihood ratio test, in contrast, automatically favors more complex models (e.g.\ ones with more parameters), because such model can be tuned to fit the data better without them suffering any penalty on account of their increased complexity: an arbitrarily complex model will fit the data arbitrarily well. The evidence automatically balances model complexity against fitting accuracy to find the most plausible model, as described in the above references.

If we had good reasons to adopt unequal model priors (i.e.\ other information favored one model over another), then we should instead look at the product of the Bayes factor with the ratio of these priors, but this is not done here.

To account for the time uncertainties in the glacial terminations, we interpret a termination time as a Gaussian measurement model
\begin{equation}
  P(t_j|\tau_j) = \frac{1}{\sqrt{2\pi}\sigma_j}e^{(t_j-\tau_j)^2/2\sigma_j^2}
  \label{eqn:measurement}
\end{equation}
where $t_j$ is the {\it measured} time of termination $j$ (identified from a stacked $\delta^{18}$O record), $\sigma_j$ is the estimated uncertainty in that measurement and $\tau_j$ is the (unknown) {\it true} termination time.

If $D$ comprises $N$ independently measured events, then the probability of observing the complete data set $D=\{t_j\}$ is just the product
\begin{equation}
  \begin{array}{r@{}l}
  P(D|\boldsymbol{\theta},M)&{}\displaystyle=\prod\limits_j^N P(t_j|\boldsymbol{\theta},M)\\
  &{}\displaystyle=\prod\limits_j^N \int_{\tau_j}P(t_j|\tau_j)P(\tau_j|\boldsymbol{\theta},M)d\tau_j
\end{array}
\label{eqn:likelihood}
\end{equation}
where the second line just follows from the marginalization rule of probability.
$P(t_j|\boldsymbol{\theta},M)$, the {\em event likelihood}, is the probability that an event (termination) $j$ is observed at time $t_j$. It is equal to the integral of the product of the measurement model with the model-predicted probability of the true time of the event, $P(\tau_j|\boldsymbol{\theta},M)$, over all values of the true time. That is, we marginalize (average) over the unknown true time. (This is explained further in section \ref{sec:termination} after we have introduced the models.)

This model-predicted probability of the times of the events, i.e.\ the deglaciations, is the {\em time series model}. This will be derived in section~\ref{sec:model} from the orbital forcing and pacing models.  

We then have all the ingredients we need to calculate the likelihood (equation~\ref{eqn:likelihood}), and therefore the evidence (equation~\ref{eqn:bayes2}) for a given time series model for a given data set.  Both the likelihood calculation and the evidence calculation involve an integral. We perform these numerically. The former is one dimensional (over time), so is straightforward.  The latter is multi-dimensional (over the model parameters), so we use a Monte Carlo method. This involves drawing parameter samples from the parameter prior distribution, $P(\boldsymbol{\theta}|M)$, calculating the likelihood for each, and then averaging the result. In each case we draw $10^5$ samples.

The Bayes factor is a positive number. The larger it is compared to unity, the more we favor model 1 over model 2.  Based on the criterion given by \cite{kass95}, we conclude that model 1 should be favored over model 2 if the Bayes factor is more than 10 (and 2 over 1 if it is less than 0.1). If the Bayes factor lies between 0.1 and 10, we cannot favor either model.

\section{Time series models}\label{sec:model}

In section \ref{sec:forcing} we introduce various climate forcing models, such as those based on variations of the Earth orbital parameters. In section \ref{sec:pacing} we define the pacing models. We use this term in a somewhat narrower sense than is often used in the literature \citep{saltzman84,tziperman06}. Here a pacing model is one which modulates the effect of a continuously variable forcing mechanism through the introduction of a threshold. Specifically, the ice volume is unaffected by the forcing mechanism until the ice volume exceeds some threshold, where the value of this threshold depends on the magnitude of the forcing. Having defined the forcing and pacing models, we use them in section \ref{sec:termination} to predict a sequence of glacial termination times.  For a given forcing/pacing model $M$, and values of its parameters $\boldsymbol{\theta}$, this is the term $P(\tau_j|\boldsymbol{\theta},M)$ in equation~\ref{eqn:likelihood}. In section \ref{sec:comparison} we will compare these model-predicted terminations with the measured ones, using the the Bayesian approach to compare the overall ability of the models to explain the data.

\subsection{Forcing models}\label{sec:forcing}

Insolation influences the climate in a number of ways, both directly through mechanisms such as heating the lower atmosphere, and indirectly through modifying the ice accumulation rate and other mechanisms \citep{berger78,Berger1978139,saltzman90}. Mainstream thinking holds that climate change is most sensitive to the northern summer insolation at high latitudes because the temperature in continental areas, of which there is more in the northern hemisphere, is critical for ice melting or sublimation \citep{milankovitch41}. The summer insolation at high latitudes depends on the geometry of the Earth's orbit and the inclination of Earth's spin axis, and thus depends on the eccentricity, precession, and obliquity (hereafter referred to collectively as ``orbital elements'', even though obliquity is not orbital).
Variations in these alter how the insolation varies with season (from orbital and axial precession), with latitude (from obliquity changes), and with time scale (e.g.\ eccentricity variations occur at dominant periods of 100\,kyr and 400\,kyr).

Milankovitch proposed that the combination of orbital elements which gives rise to the measured summer insolation at $65^{\circ}$N is crucial to generating the glacial-interglacial cycles \citep{milankovitch41,hays76}. To model orbital forcings more generally, we define an orbital forcing model, $f(t)$, as a combination of eccentricity, precession, and obliquity, which is proportional to the insolation over certain time scales, seasons, and latitudes. 
We build the following forcing models based on the reconstructed time-varying eccentricity, $f_{\rm E}(t)$, precession, $f_{\rm P}(t)$, obliquity, $f_{\rm T}(t)$, and four different combinations thereof:
\begin{equation}
\begin{aligned}
f_{\rm E}(t) \,&=\, e(t)\\
f_{\rm P}(t) \,&=\, e(t) \sin(\omega(t)-\phi)\\
f_{\rm T}(t) \,&=\, \epsilon(t)\\
f_{\rm EP}(t) \,&=\, \alpha^{1/2}f_{\rm E}(t) + (1-\alpha)^{1/2}f_{\rm P}(t)\\
f_{\rm ET}(t) \,&=\, \alpha^{1/2}f_{\rm E}(t) + (1-\alpha)^{1/2}f_{\rm T}(t)\\
f_{\rm PT}(t) \,&=\, \alpha^{1/2}f_{\rm P}(t) + (1-\alpha)^{1/2}f_{\rm T}(t)\\
f_{\rm EPT}(t) \,&=\, \alpha^{1/2}f_{\rm E}(t) + \beta^{1/2}f_{\rm P}(t) + (1-\alpha-\beta)^{1/2}f_{\rm T}(t),
\label{eqn:ts_function}
\end{aligned}
\end{equation}
where $e(t)$, $\epsilon(t)$, and $e(t)\sin(\omega(t)-\phi)$ are the eccentricity, obliquity, and precession index (or climatic precession), respectively. $\omega(t)$ is the angle between perihelion and the vernal equinox, and $\phi$ is a parameter controlling the phase of the precession. We use the variations of these three orbital elements over the past 2\,Myr as calculated by \cite{laskar04}. We standardize each of $f_E(t)$, $f_P(t)$, and $f_T(t)$ to have zero mean and unit variance, and then combine them to generate the compound models. $\alpha$ and $\beta$ are contribution factors which determine the relative contribution of each component in the compound models, where $0\leq\alpha\leq 1$ and $0\leq\beta\leq 1$. In addition to these models, we also use the daily-averaged insolation at 65$^\circ$N on July 21 as a proxy for the Milankovitch forcing, $f_{\rm CMF}$. 

Beyond orbital forcings, we also consider the influence of variations of the Earth's orbital inclination and of  the cosmic ray flux. To do this we build an inclination-based forcing model, $f_{\rm Inc}(t)$, using the orbital inclination calculated by \cite{muller97}, and we model the cosmic ray forcing as a geomagnetic paleointensity (GPI) time series (standardized to the mean and unit variance), $f_G(t)$, as collected by \cite{channell09}.

All forcing models and corresponding prior distributions over their parameters (``forcing parameters'') are shown in Table \ref{tab:ts_models}. In this table and the following sections, all parameters are treated as dimensionless variables by setting the time unit to be 1\,kyr (ice volume is on a relative scale). For the precession model, we set $\phi=0$ to treat precession according to the Milankovitch theory (although in section \ref{sec:sensitivity} we will allow the phase of the precession to vary in order to check the sensitivity of our results to this assumption). As we do not have any prior information about the values of the contribution factors in the compound models, we adopt uniform prior distributions over the interval $[0,1]$ for these.

\begin{table*}[t]
  \centering
  \caption{The termination models and corresponding forcing models and parameters. In addition to any forcing model parameters listed, the termination models have pacing parameters and the background fraction parameter. The prior distributions of these parameters are described in sections \ref{sec:forcing}, \ref{sec:pacing}, and \ref{sec:termination}, respectively. }
  \label{tab:ts_models}
  \begin{tabular}{llll}
  \hline
  Termination & Description & Forcing & Forcing model \\
  model         &                   & model & parameters\\
  \hline
    Periodic &100\,kyr pure periodic model & None & --- \\
    Eccentricity &Eccentricity& $f_{\rm E}(t)$ & --- \\
    Precession &Precession& $f_{\rm P}(t)$ & $\phi$\\
    Tilt &Tilt or obliquity& $f_{\rm T}(t)$ & ---\\
    EP &Eccentricity plus Precession& $f_{\rm EP}(t)$ & $\alpha$, $\phi$\\
    ET &Eccentricity plus Tilt& $f_{\rm ET}(t)$ &$\alpha$\\
    PT &Precession plus Tilt& $f_{\rm PT}(t)$ &$\alpha$, $\phi$\\
    EPT &Eccentricity plus Precession plus Tilt&$f_{\rm EPT}(t)$&$\alpha$, $\beta$, $\phi$\\
    CMF &(Classical) Milankovitch forcing& $f_{\rm CMF}(t)$ &---\\
    Inclination &Inclination& $f_{\rm Inc}(t)$ &---\\
    GPI &Geomagnetic paleointensity& $f_{\rm G}(t)$ &---\\
    \hline
  \end{tabular}
\end{table*}

\begin{figure*}[ht!]
  \centering
  \includegraphics[scale=0.7]{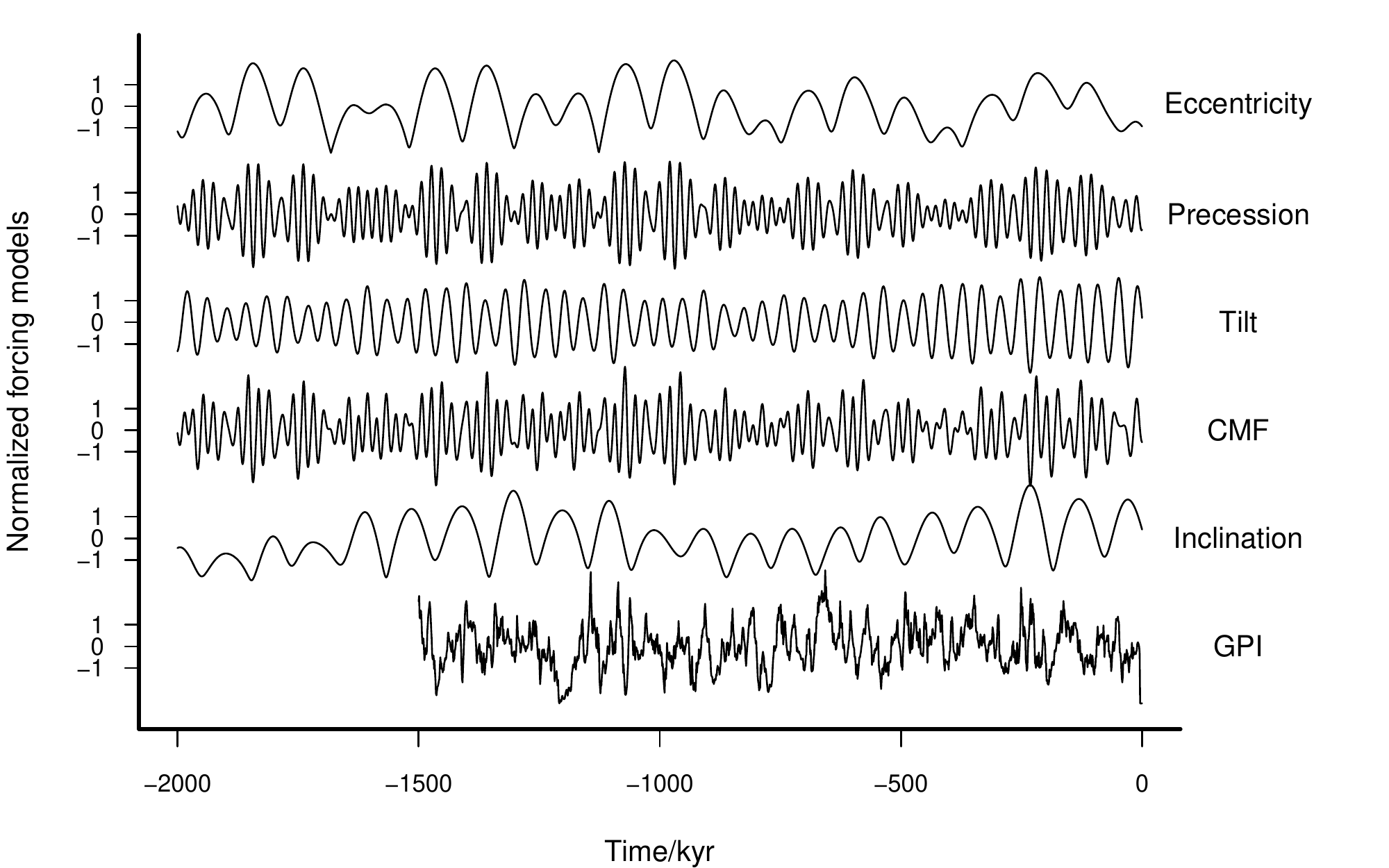}
  \caption{The single-component forcing models.
    A deglaciation is likely to be triggered by a peak in the forcing. The values of eccentricity, precession, obliquity and Milankovitch forcing (CMF) are calculated by \cite{laskar04}, the orbital inclination relative to the invariable plane is given by \cite{muller97}, and the GPI record is from \cite{channell09}.}
  \label{fig:forcing_models}
  \end{figure*}

Figure \ref{fig:forcing_models} shows the single-component forcing models (which do not have any adjustable parameters). All forcing models will be included in pacing models and corresponding termination models in the following sections. Hereafter, for each forcing model, the corresponding pacing and termination models share the same name as shown in the first column of Table \ref{tab:ts_models}. 

\subsection{Pacing models}\label{sec:pacing}

\begin{figure*}[ht!]
  \centering
  \includegraphics[scale=0.8]{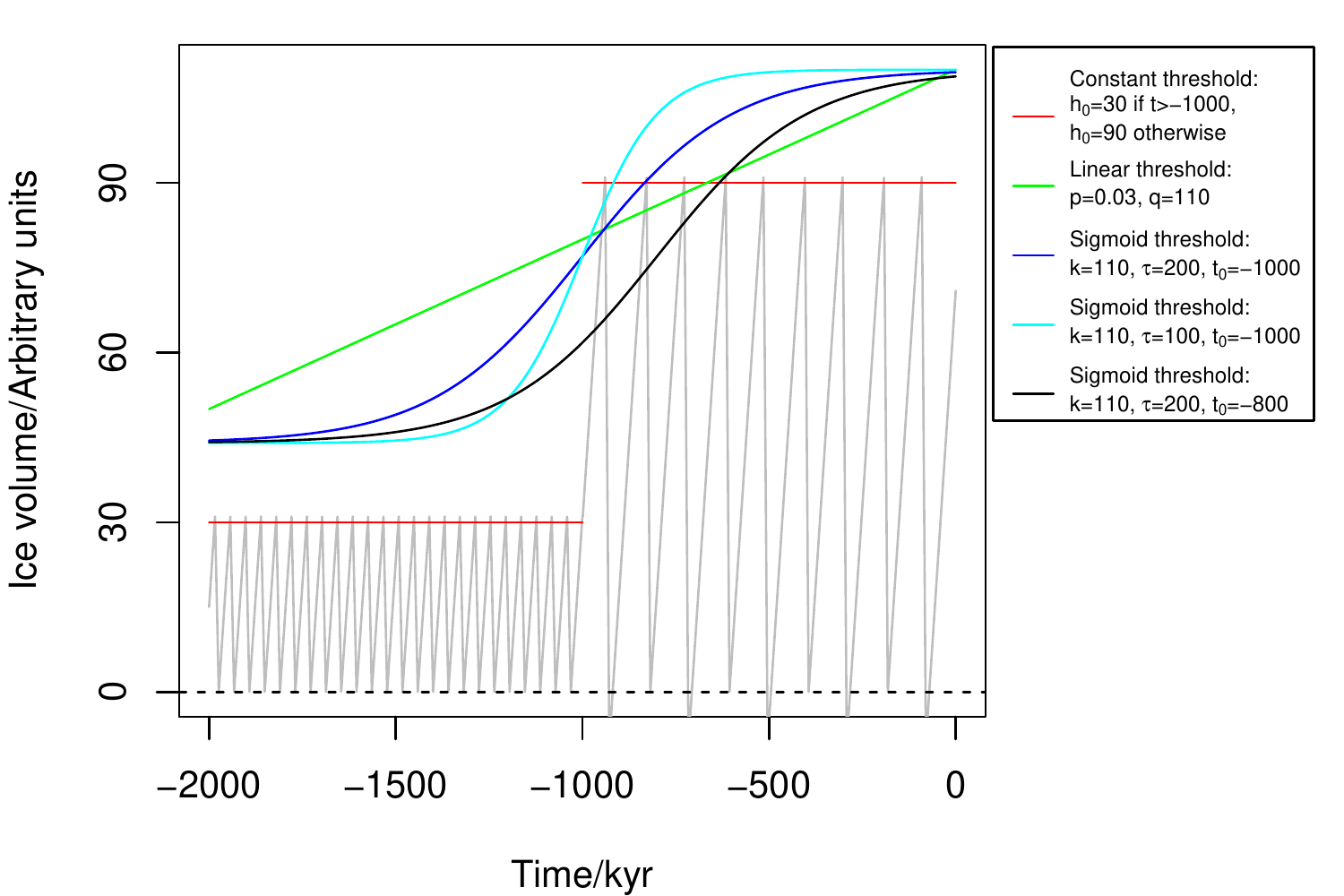}
  \caption{Effect of the threshold in the pacing model. Different values of the threshold, $h_0(t)$, are shown: constant (red), linear (green), sigmoidal (blue, cyan, black). The legend shows the values of the parameters of the linear and sigmoid background thresholds according to equations \ref{eqn:linear} and \ref{eqn:sigmoid} respectively. The Periodic model is achieved using a constant threshold over some time span.  By changing it from $h_0=30$ in the early Pleistocene to $h_0=90$ in the late Pleistocene, we can reproduce an abrupt change in the period of ice volume variations from $\sim$41\,kyr to $\sim$100\,kyr.}
  \label{fig:icevolume}
\end{figure*}

As described earlier, we use the term ``pacing'' to mean that some aspect of the climate system is independent of external forcings until the climate system reaches a threshold, whereby the value of this threshold is dependent upon the forcing. We model the pacing effect on ice volume variations using the deterministic version of the stochastic model introduced by \cite{huybers05}. In that model the ice volume at time $t$ is
\begin{equation}
  v(t)=v(t-\Delta t)+\eta(t) \quad \quad \text{and if } v(t)>h(t) \text{ then terminate},
\label{eqn:deterministic}
\end{equation}
where
\begin{equation}
  h(t)=h_0-af(t),
  \label{eqn:threshold}
\end{equation}
and $\Delta t$ is a constant time interval. Thus the ice volume changes in discrete steps until it passes a threshold $h(t)$, which is itself modulated by a climate forcing $f(t)$ with a contribution factor $a$. The initial ice volume is $v_0$ and the {\em background threshold}, $h_0$, is either a constant or can itself vary with time. We set $\eta(t)$ to be unity while the threshold has not been reached; after that the glaciation is terminated by setting $\eta(t)$ to a constant negative value such that the ice volume linearly decreases to 0 within 10\,kyr of the threshold having been exceeded.\footnote{In practice the ice volume can go slightly negative due to the finite value of $\Delta t$, but this is of no practical consequence.} After this $\eta(t)$ is set to unity, the next cycle starts.
The threshold and the deglaciation duration are chosen to generate approximately 100 and 41\,kyr glacial cycles \citep{huybers05}. 
If the contribution factor $a$ is zero, the ice volume will vary with a period modulated by the background threshold, $h_0$. We define this model as the Periodic model. In general the period may vary with time.
However, if
$h_0$ is constant, then the Periodic model has a constant period of value $h_0+10$\,kyr. Because $h_0$ controls the period of ice volume variations, different values of $h_0$ are required to model the 100\,kyr cycles in the late Pleistocene and the 41\,kyr cycles in the early Pleistocene (see Figure \ref{fig:icevolume}).  We therefore first build pacing models to separately predict the deglaciations over the early and late Pleistocene using the constant background threshold model. We then use a varying background threshold (either linear or sigmoidal) to try to model the whole Pleistocene. We now describe these models in more detail.

\subsubsection{Constant background threshold}\label{sec:pacing_constant}

\begin{figure*}[ht!]
  \centering
  \includegraphics[scale=0.6]{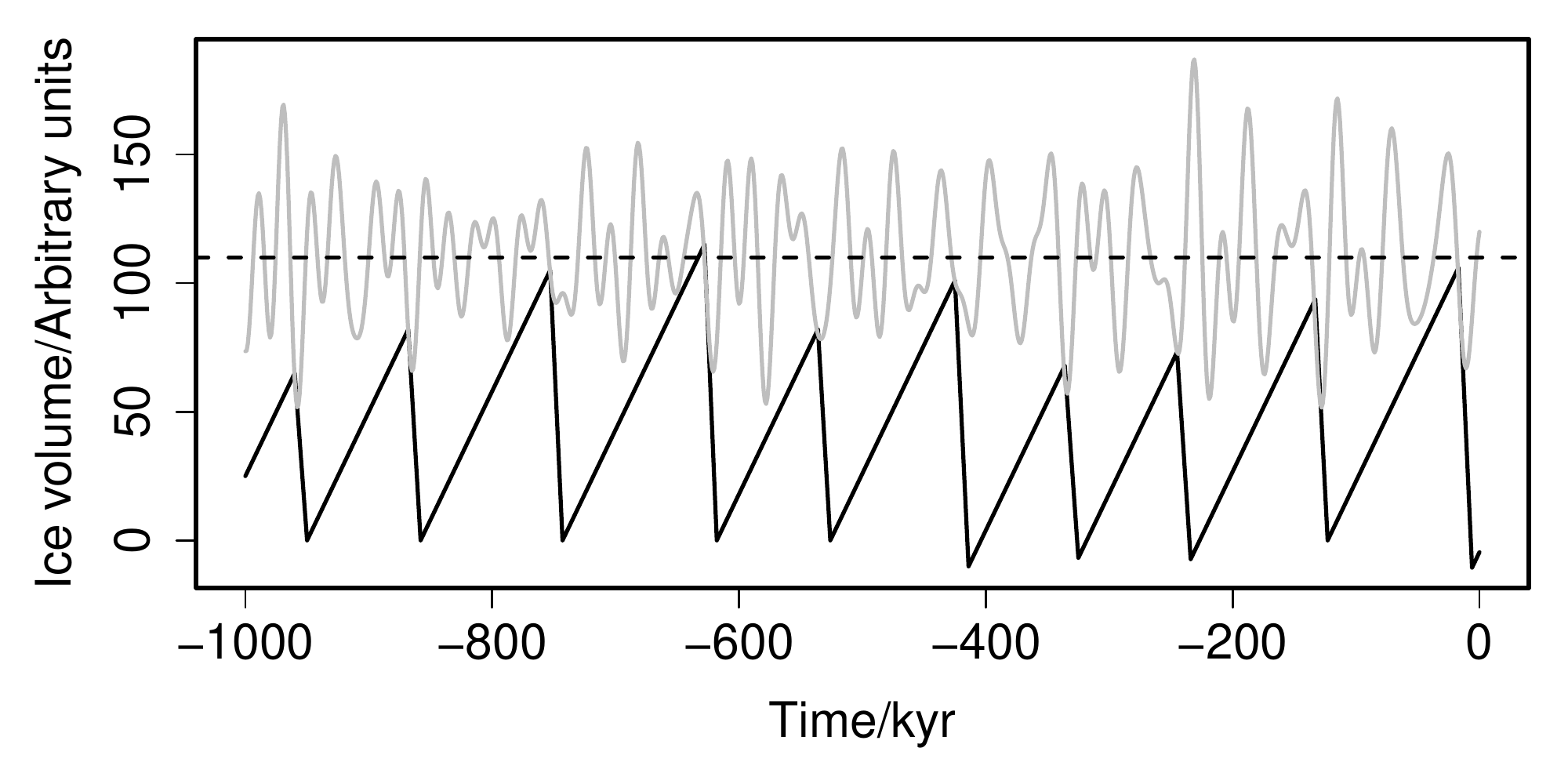}
  \caption{A pacing model with threshold $h(t)$ modulated by the PT forcing model with $\alpha=0.5$ and $\phi=0$ (equation \ref{eqn:ts_function}). The pacing model parameters are: background threshold $h_0=90$; initial ice volume $v_0=25$; contribution factor of forcing $a=25$. The dashed line denotes the constant threshold, and the grey line represents the threshold modulated by the PT forcing model, i.e. $h(t)=h_0-af_{\rm PT}(t;\alpha=0.5,\phi=0)$. }
  \label{fig:pacing_model}
\end{figure*}

A constant background threshold is appropriate for modeling glacial-interglacial cycles without a transition such as the MPT.  One realization of such a pacing model with the threshold modulated by a PT forcing model is shown in Figure \ref{fig:pacing_model}. The ice volume grows until it passes the forcing-modulated threshold. The ice volume then decreases rapidly to zero within the next 10\,kyr. We see that a deglaciation tends to occur when the insolation is near a local maximum. Hence the pacing model (equations \ref{eqn:deterministic} and \ref{eqn:threshold}) can generate $\sim$100\,kyr saw-tooth cycles which enables a forcing mechanism to pace the phase of these cycles. 

The pacing model has three parameters: $v_0$, $h_0$, $a$.  Rather than fixing these to some expected values, we assign a probability distribution to them.  This is the prior which appears in equation~\ref{eqn:bayes2}, which shows that by averaging the likelihood over values drawn from this prior we get the evidence for the model.

As described above, a periodic pacing model is generated by adopting a constant threshold, $h(t)=h_0$ and $a=0$. When forcings are added onto the constant threshold (to give $a\neq0$), the ice volume variations then have an average period of about $(h_0+10-a)$\,kyr, because ice volume accumulation tends to terminate at a forcing maxima. For this reason we use different prior distributions on $a$ and $h_0$ depending on whether we
are trying to model the early (41\,kyr cycles) or late (100\,kyr cycles) Pleistocene. 
Specifically, we use prior distributions for $v_0$, $h_0$, and $a$ which are uniform over the following intervals (and zero outside): $0<v_0<90\gamma$, $90\gamma<h_0<130\gamma$, $15\gamma<a<35\gamma$, where $\gamma=0.4$ when we model $\sim$41\,kyr cycles and $\gamma=1$ when we model $\sim$100\,kyr cycles.
The range of $v_0$ is just the range of the ice volume variation, while the mean values of the prior distributions of $h_0$ and $a$ with $\gamma=1$ are the fitted values obtained by \cite{huybers11}. For the periodic model, $a$ is zero and $h_0$ has a uniform prior distribution over $70\gamma <h_0< 110\gamma$. In section \ref{sec:sensitivity}, we will check how sensitive our results are to this choice of priors.

\subsubsection{Linear trend background threshold}\label{sec:pacing_linear}

The constant background threshold model is incapable of modeling the transition from the 41\,kyr world to the 100\,kyr world. If we treat $h_0$ as a step function as shown in Figure \ref{fig:icevolume} (red lines), the corresponding pacing model predicts an abrupt MPT with an extra parameter (the time of the transition). But to model the MPT, we will introduce another two versions of the pacing model by allowing the background threshold to vary with time (linearly and nonlinearly).

Studies have suggested various mechanisms which may be involved in climate change before and after the MPT \citep{saltzman84,maasch90,ghil94,raymo97,paillard98,clark99,tziperman03,ashkenazy04}. H07 suggests that a simple model with a threshold modulated by obliquity and a linear trend can explain changes in glacial variability over the last 2\,Myr without invoking complex mechanisms. To investigate this, we replace the threshold constant $h_0$ with a linear trend in time
\begin{equation}
  h_0=pt+q,
  \label{eqn:linear}
\end{equation}
where $p$ and $q$ are the slope and intercept of the trend respectively. To predict the transition from 41\,kyr cycles to 100\,kyr cycles with reasonable parameter sets, we adopt the following uniform prior distributions for the pacing parameters: $0<v_0<36$, $0<p<0.1$, $106<q<146$ and $10<a<30$. For the Periodic model we use $a=0$ and a uniform prior for $q$ between 86 and 126. These ranges are adopted so that the pacing model predicts the 41\,kyr and 100\,kyr cycles with similar period uncertainties as produced by the ranges of parameters in the pacing model with a constant background threshold (section \ref{sec:pacing_constant}).

An example of the linear trend is shown with the green line in Figure \ref{fig:icevolume}. If the threshold is not modulated by any forcing (i.e.\ $a=0$, the Periodic model), then the pacing model generates a gradual transition from 50\,kyr cycles 2\,Ma to 110\,kyr cycles at the present. 

\subsubsection{Sigmoid trend background threshold}\label{sec:pacing_sigmoid}

To enable a more rapid onset of the MPT, we introduce another version of the pacing model with a nonlinear trend in the background threshold, defined using the sigmoid function as
\begin{equation}
  h_0=0.6k/(1+e^{-(t-t_0)/\tau})+0.4k,
  \label{eqn:sigmoid}
\end{equation}
where $k$ is a scaling factor, $t_0$ denotes the transition time, and $\tau$ represents the time scale of the MPT. The uniform priors of the parameters of this version of pacing models are set to be: $0<v_0< 36$, $90<k<130$, $10<\tau<500$, $10<a<30$, and $-700<t_0<-1250$, as motivated by the range of MPT time given by \cite{clark06}. For the Periodic model we set $a=0$ and change the range of $k$ to be $70<k<110$. The reason for choosing these priors is the same as given in section \ref{sec:pacing_linear}.

Figure \ref{fig:icevolume} illustrates this model.  A late transition time, $t_0$, moves the trend to the present, and a smaller transition time scale, $\tau$, generates a more rapid transition. 
The values of $0.6k$ and $0.4k$ in the above equation are set in order to rescale the trend model such that the ice volume threshold including a sigmoid trend allows both $\sim$41\,kyr and $\sim$100\,kyr ice volume variations.

\subsection{Termination models}\label{sec:termination}

Using the same method described in section \ref{sec:data} for the data, we identify glacial terminations in the ice volume time series generated by the pacing models. The age uncertainty of each termination is equal to half of the duration of the termination. As with the data, a single termination is represented as a Gaussian probability distribution over time, which is just the term  $P(\tau_j|\boldsymbol{\theta},M)$ in equation \ref{eqn:likelihood} (see section \ref{sec:bayes}). The full set of predicted terminations forms the time series model which we will compare with the data. We use the term ``termination model'' to refer to the combination of a forcing model and a pacing model, which together has a number of parameters. These are listed in Table \ref{tab:ts_models}. Each of these termination models can have different background threshold models, as was explained in section \ref{sec:pacing}.

\begin{figure*}[ht!]
  \centering
  \includegraphics[scale=0.6]{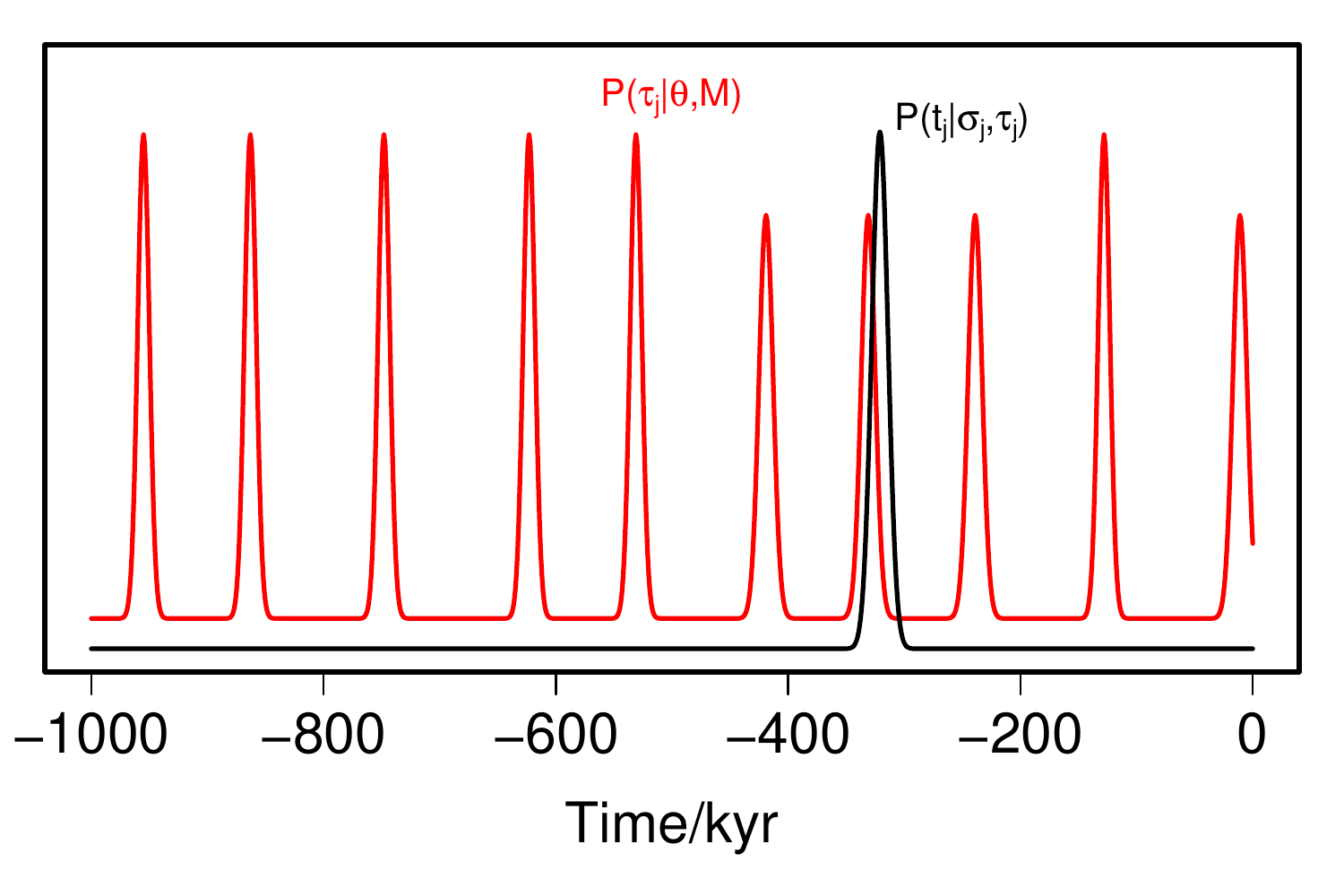}
\caption{Schematic illustration of the components in the likelihood calculation (equation \ref{eqn:likelihood}). The red line is the termination model generated from the pacing model shown in Figure \ref{fig:pacing_model}. The black line represents the measured data on termination $j$. Its time and uncertainty are interpreted probabilistically as a Gaussian distribution over time.}
  \label{fig:termination_model}
\end{figure*}

Figure \ref{fig:termination_model} shows schematically how we compare the model-derived terminations (red line) with the data on one termination (black line).  The event likelihood (the integral in equation \ref{eqn:likelihood}) for a termination is calculated by integrating over time the product of the probability distribution of the observed time of the termination, $P(t_j|\sigma_j,\tau_j)$, with the model prediction of the true termination time, $P(\tau_j|\boldsymbol{\theta},M)$. The product of event likelihoods for all terminations in a data set is the likelihood for the termination model with specific values of the parameters of the forcing and pacing model.
By calculating the likelihood for many different values of those parameters (drawn from their prior distributions), and averaging them, we arrive at the evidence for that termination model (equation~\ref{eqn:bayes2}).

To accommodate other contributions from the climate system to the timing of a termination, we add a constant background probability to the termination model. This is defined using the background fraction $b=H_b/(H_b+H_g)$, where $H_b$ is the amplitude of the background and $H_g$ is the difference between the maximum and minimum of the Gaussian sequence. 
The background fraction is a parameter of the model which we do not measure, so we assign it a prior (uniform from 0 to 0.1) and marginalize over this too. 

Let us summarize our modelling procedure. A forcing model (Figure \ref{fig:forcing_models}) modulates the ice volume threshold (equation \ref{eqn:threshold}) of the pacing model (equation \ref{eqn:deterministic}) from which the
termination model (e.g.\ red line in Figure \ref{fig:termination_model}) is derived. This is then compared with a sequence of terminations identified from a $\delta^{18}$O data set using our Bayesian procedure.

\section{Results of the model comparison}\label{sec:comparison}

\subsection{Evidence and Bayes factor}\label{sec:BF}

We calculate the Bayesian evidence of the termination models listed in Table \ref{tab:ts_models} for each of the data sets shown in Table \ref{tab:terminations}. 
We calculate this for terminations extending over three different time spans: 1\,Ma to 0\,Ma, 2\,Ma to 1\,Ma and 2\,Ma to 0\,Ma.  The first time span is the same as that chosen by \cite{huybers11}. However, other studies claim that the onset of 100\,kyr cycles occurred around 0.8\,Ma. We will examine in section \ref{sec:sensitivity} how sensitive our results are to the choice of time span. According to the time span in question, we need to choose the appropriate pacing model, because this determines the dominant period.

The Bayes factor (BF) is just the ratio of the evidence for two models. Rather than reporting Bayes factors for various pairs of models, we will report them for all models relative to a simple reference termination model. This reference model is just a uniform probability distribution over the time of deglaciations, and has no parameters. It corresponds to a constant probability in time of a deglaciation, but its choice is arbitrary as it just serves to put the evidences on a convenient scale.

Bayes factors should only be used to compare different models for a common data set. 
This is because their definition requires that the factor $P(D)$ in equation~\ref{eqn:bayes3} cancels out.

\subsubsection{Late Pleistocene (1-0\,Ma)}\label{sec:LP}

The deglaciations identified using H07's method (in the data sets HA, HB, HP, LR04, and LRH) contain many minor terminations which may be better explained by models which predict $\sim$41\,kyr cycles. Thus, we choose constant background thresholds with $\gamma=1$ and $\gamma=0.4$ for all termination models in order to predict 100\,kyr and 41\,kyr variations, respectively, over the past 1\,Myr. 

\begin{figure*}
    \centering
    \vspace{-1in}
  \includegraphics[scale=0.4]{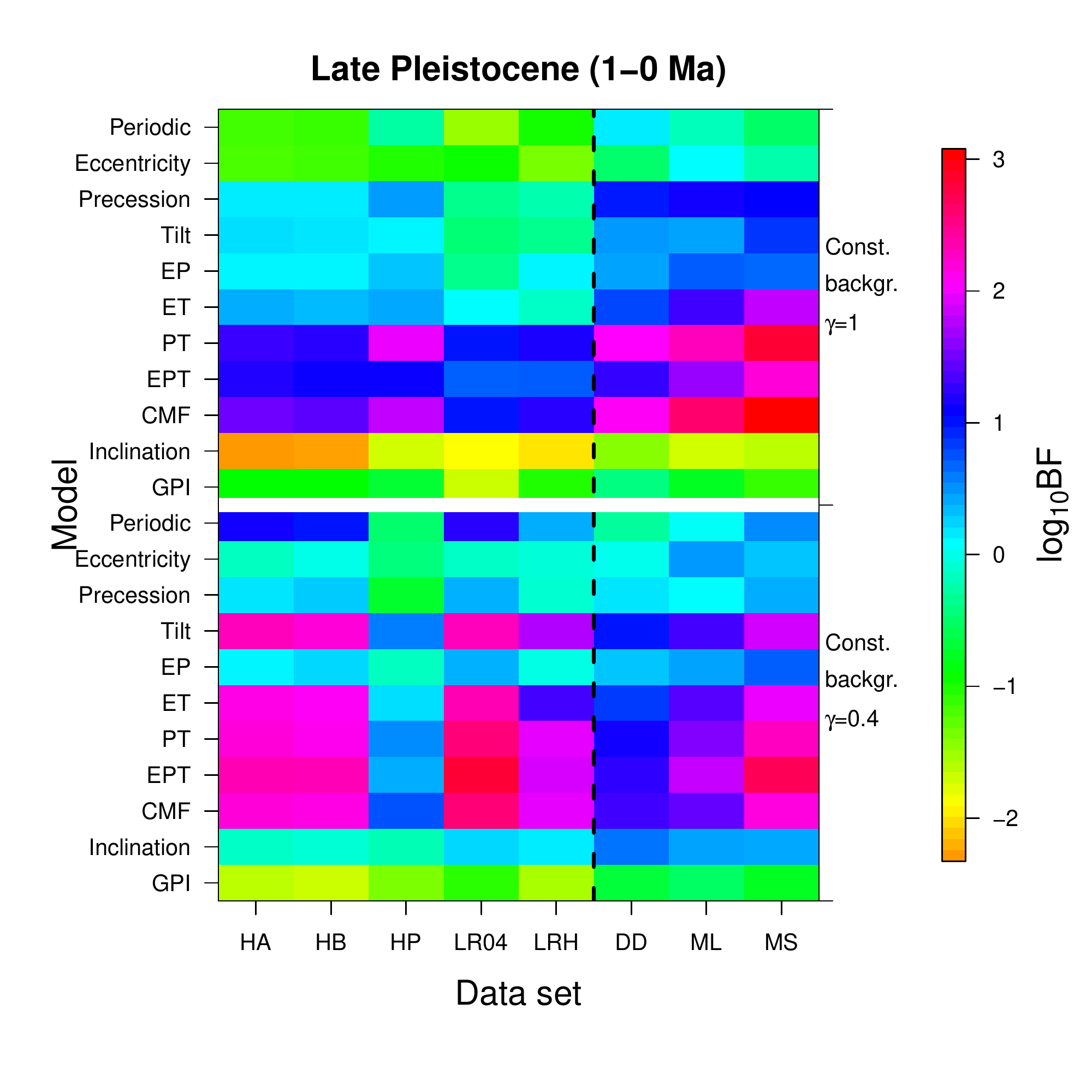}
  \includegraphics[scale=0.4]{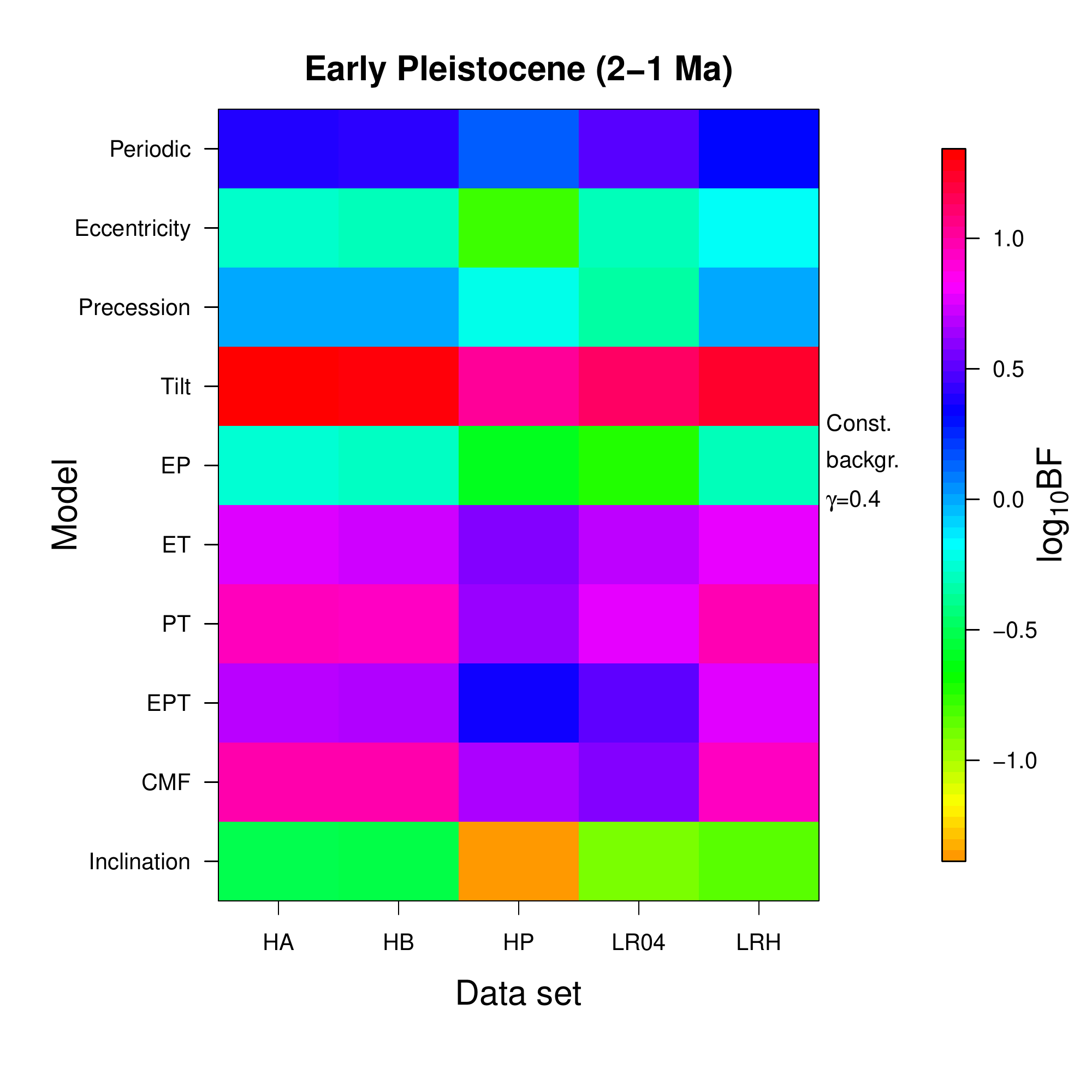}
  \includegraphics[scale=0.4]{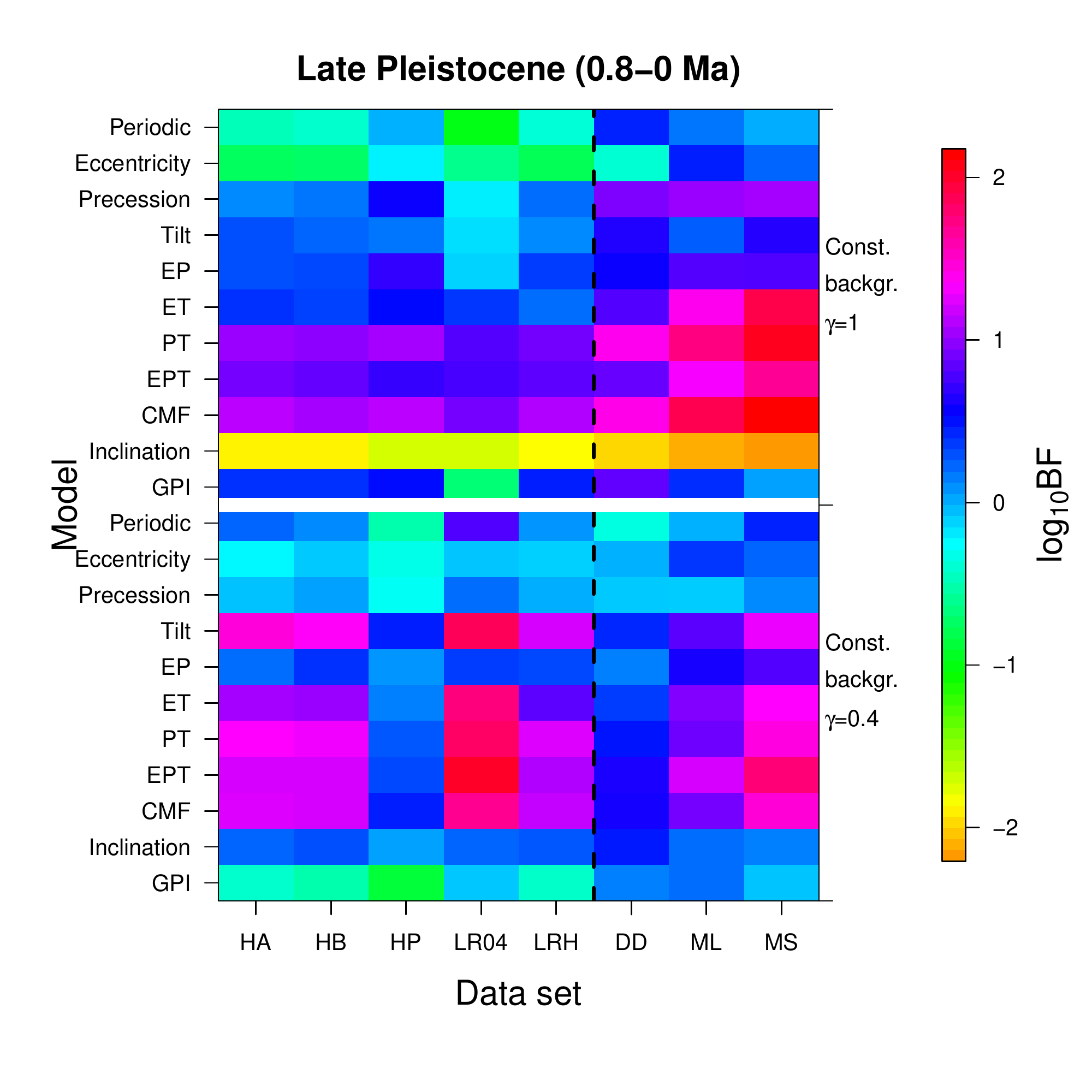}
  \includegraphics[scale=0.4]{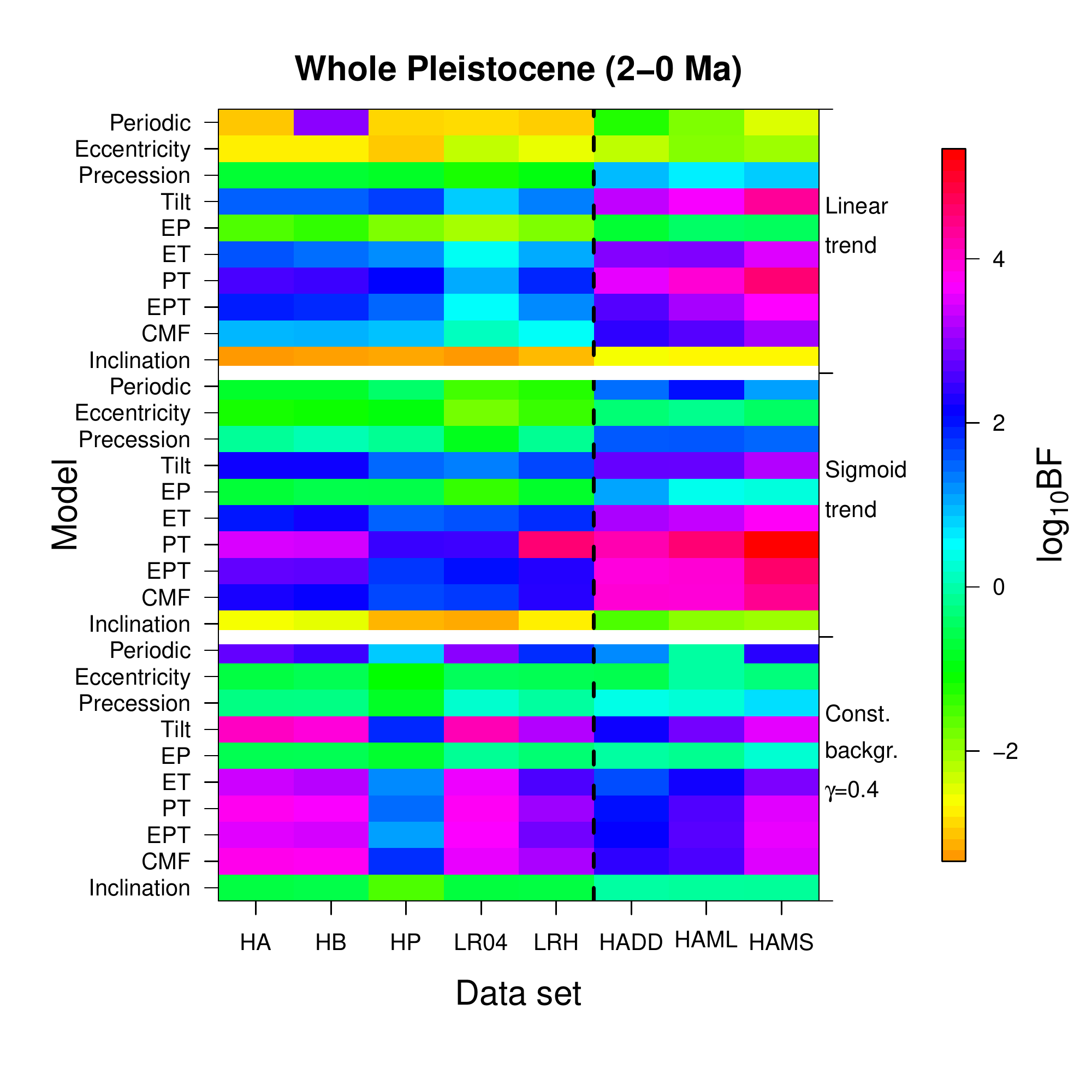}
  \caption{The Bayes factors relative to the uniform model for terminations occurring over the past 1\,Myr (upper left), from 2\,Ma to 1\,Ma (upper right), over the past 0.8\,Myr (lower left), and over the past 2\,Myr (lower right).  The logarithm of the Bayes factor is shown on a color scale for each model (vertical axis) and data set (horizontal axis).  Upper left panel: the models above and below the white line have constant background threshold defined by $\gamma=1$ and $\gamma=0.4$, respectively. Upper right panel: all models have a constant background threshold defined by $\gamma=0.4$. Lower left panel: Same as the upper left panel but for terminations over the past 0.8\,Myr. Lower right panel: the upper, middle, and lower blocks (of ten models each, separated by the white line) use a linear trend, a sigmoid trend, and a constant background threshold (respectively) with $\gamma=0.4$. In all panels except the top right one, the data sets on the left side of the dashed line include minor late-Pleistocene terminations while the data sets on the right side do not.
}
  \label{fig:BF_2D}
\end{figure*}

The BF for each termination model relative to the uniform model is shown in Figure \ref{fig:BF_2D}. We see that the HA, HB, LR04, and LRH data sets favor the models with a tilt component and with $\gamma=0.4$. Although compound models such as EPT and CMF sometimes have BFs slightly higher than the Tilt model, precession and eccentricity may not be necessary to explain the terminations identified in these data sets. 

The HP data set favors the PT model with $\gamma=1$. This could be caused by a mismatch between the terminations identified in HP and the terminations identified in other data sets. For example, around the time of termination 6 (Figure \ref{fig:huybers_data}), two terminations are identified in HP while only one termination is identified in other data sets. The discrepancy between HP and other data sets is larger before 0.8\,Ma, which indicates a more ambiguous definition of terminations, particularly for planktic $\delta^{18}$O.  On account of this, in section \ref{sec:sensitivity} we will narrow the time span to 0-0.8\,Ma (a more conservative time scale of late Pleistocene). Nevertheless, for all the data sets containing minor terminations, tilt is a common factor in the preferred models.  

For the DD, ML, and MS data sets, the PT and CMF models with $\gamma=1$ are favored. In other words, 
the major terminations are better predicted by a model involving precession and tilt rather than either alone, although tilt alone can pace minor terminations. Because the EPT and CMF models have lower BFs than the PT model, the eccentricity component is unlikely to pace the glacial terminations directly. Yet eccentricity can determine the glacial terminations indirectly through modulating the amplitude of the precession maxima (i.e.\ $e\sin{\omega}$). A similar conclusion was drawn by \cite{huybers11} using p-values. 
We note that the rejection of a null hypothesis in this way does not automatically validate the alternative  hypothesis.  The Bayesian approach allows one to directly compare multiple models in a symmetric fashion.

Since the late Pleistocene is characterized predominantly by major terminations, we conclude that late Pleistocene climate change is paced by a combination of obliquity and precession. This does not automatically imply that there is no link between major terminations and eccentricity variations. Eccentricity may determine the 100 kyr cycles in the late Pleistocene, while obliquity and precession influence the exact timing of the terminations \citep{lisiecki10}.  This could be studied in future work by introducing an eccentricity dependence into the pacing model.

\subsubsection{Early Pleistocene (2-1\,Ma)}\label{sec:EP}

Here we only consider the HA, HP, HB, L04, and LRH data sets, because the DD, ML, and MS sets have no terminations in the early Pleistocene.  We only calculate BFs for models with $\gamma=0.4$ (and not $\gamma=1$), because this reproduces periods on the order of 41\,kyr, and such cycles are obvious in all data sets (Figure \ref{fig:huybers_data}).  We exclude the GPI model because the GPI record has a time span less than 2\,Myr. The BFs for the termination models are shown in the upper right panel of Figure \ref{fig:BF_2D}.

We see that the Tilt model is favored by all data sets. The combination of tilt with other orbital elements does not give a higher BF, so we conclude that the other orbital elements do not play a major role in pacing the deglaciations over the early Pleistocene. 
It is important to realise that although the Bayesian evidence generally penalizes more complex models, this does not automatically result in a lower Bayes factor for such models. They can achieve higher Bayes factors if the model is supported by the data sufficiently strongly (see the references in section \ref{sec:bayes}).

\subsubsection{Whole Pleistocene (2-0\,Ma)}\label{sec:WP}

For the whole Pleistocene we use the data sets HA, HB, HP, LR04, and LRH as well as the hybrid data sets, HADD, HAML, and HAMS. We use pacing models with and without a trend threshold to model the terminations. The BFs for the above models and data sets are shown in the lower left panel of Figure \ref{fig:BF_2D}.

For the HA, HB, and LR04 data sets, the Tilt model with $\gamma=0.4$ is favored. Other combinations with the tilt component and with $\gamma=0.4$ yield similar BFs. However, for the HP and LRH data sets, the PT model with a sigmoid trend is favored and this model also gives high BFs for the HA, HB, and LR04 data sets. For all these data sets, the Precession, Eccentricity, and Periodic models have rather low BFs. These results indicate a major role for tilt and a minor role for precession in pacing major and minor Pleistocene deglaciations.
For all the above data sets, the CMF model with $\gamma=0.4$ has a high BF, but not higher than other models with a tilt component. CMF is an optimized version of the EPT model. Faced with different models which give similar Bayes factors, we will normally want to choose the simplest, which here is PT. We will investigate this further in section \ref{sec:sensitivity}. 

For the HADD, HAML, and HAMS data sets, the PT model with a threshold modulated by a sigmoid trend is favored, and those compound models with a tilt component also have high BFs. The whole Pleistocene deglaciations appear to be paced by the combination of precession and obliquity. This is consistent with the results for the late-Pleistocene deglaciations. 
The physical reason why precession becomes important after the MPT is beyond the scope of our work and is still under debate.

On account of the existence of the MPT, modeling the whole Pleistocene with a constant background threshold model makes little sense, so those corresponding results should not be given much weight. (This corresponds to assigning all those models a smaller model prior, $P(M)$.)  More appropriate are the models with linear and sigmoid background thresholds. Among these, we see that the EPT and CMF models have BFs about ten times lower than the PT model.
We conclude that eccentricity does not play a significant role in pacing terminations over the whole Pleistocene. We also find that the PT model with a sigmoid background threshold is more favored than the PT model with a linear background threshold, which indicates that the MPT may not be as gradual as claimed by \citep{huybers07}. We will discuss this further in section \ref{sec:sensitivity}.

According to Figure \ref{fig:BF_2D}, the Inclination and GPI models are not favored, and in fact are less favored than the reference uniform model (as BF$<$1).
Thus we find that the geomagnetic paleointensity does not pace glacial cycles over the last 2\,Myr, although we note that there is some controversy over the link between the GPI and climate change \citep{courtillot07,pierrehumbert08,bard08,courtillot08}. In contrast to the conclusion of \cite{muller97}, there we find no evidence for a link between the orbital inclination and ice volume change.

\subsection{Discrimination power}\label{sec:discrimination}

To validate our method as an effective inference tool to select out the true model, we generate simulated data from each model and then evaluate the Bayes factors for all models on these data.
The data are simulated with the following parameters for all models except the Periodic one: $h_0=110\gamma$, $a=25\gamma$, $b=0$, and $v_0=45\gamma$, where $\gamma=1$ for terminations simulated over the last 1\,Myr and $\gamma=0.4$ for the time range 2 to 1\,Ma. 
For the Periodic model we use instead of $h_0=90\gamma$ and of course $a=0$. Recall that the period of the resulting time series is approximately $h_0+10-a$. Other parameters in corresponding forcing models are fixed at $\alpha=0.5$ for compound models with two components, $\alpha=0.3$ and $\beta=0.2$ for the EPT model, and $\phi=0$ for models with a precession component.

\begin{figure*}[ht!]
  \centering
  \includegraphics[scale=0.4]{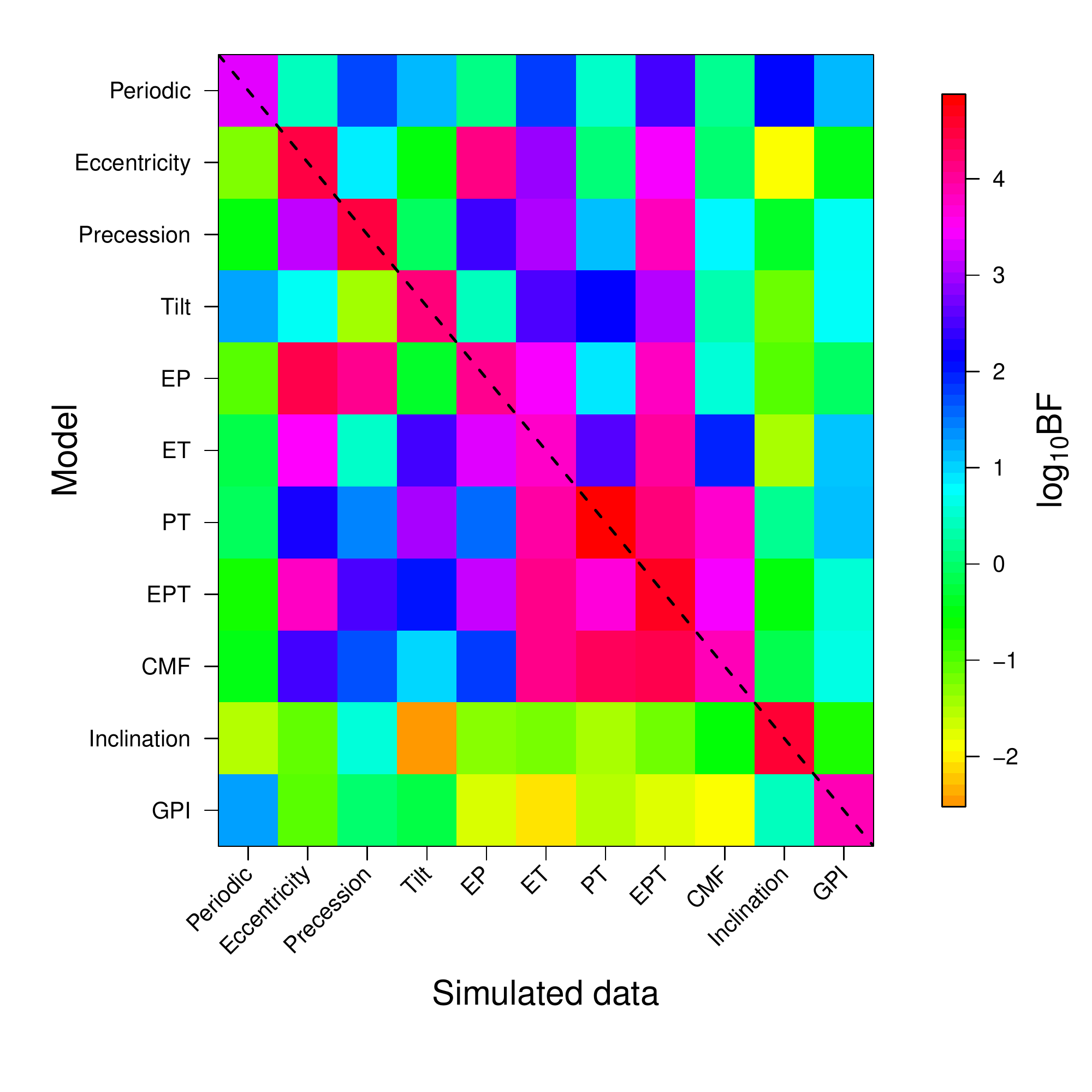}
  \includegraphics[scale=0.4]{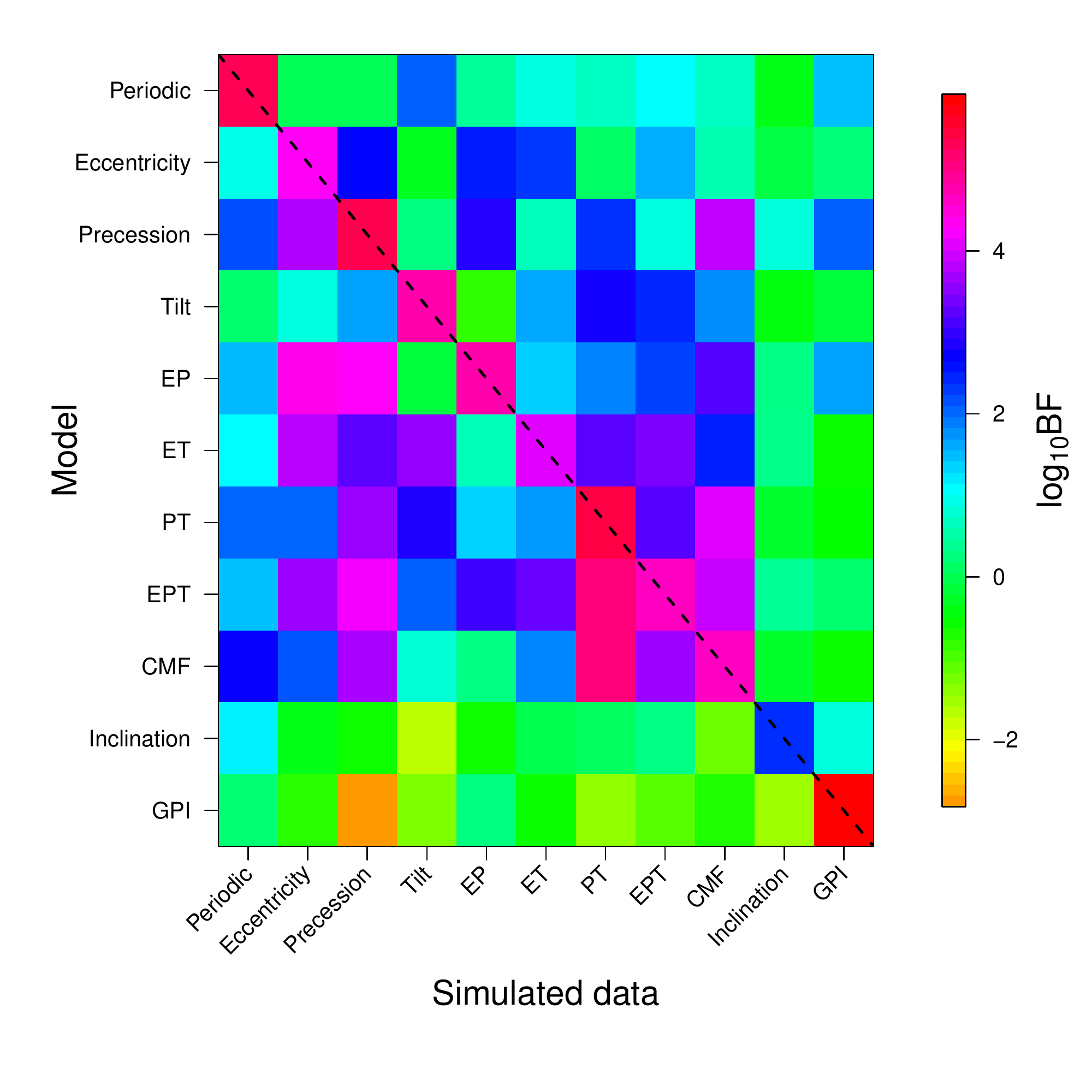}
  \caption{Discrimination power. As Figure \ref{fig:BF_2D} but for simulated data extending over the last 1 Myr (left) and from 2\,Ma to 1\,Ma (right). The dashed line indicates the results for the true model for each data set. Ideally this BF would be much higher than all other BFs in that column.}
  \label{fig:BF_2D_sim}
\end{figure*}

The BFs for simulated data over the last 1\,Myr are shown in the left panel of Figure \ref{fig:BF_2D_sim}.
We see that all models based on a single orbital element are correctly selected, although those models combining the correct single orbital element with other elements may also give comparable BFs. When models have similar BFs we would generally want to favor the one with fewest components. 
This again corresponds to using a larger value of the model prior, $P(M)$ (see section~\ref{sec:bayes}). 

Incorrect models, in contrast, generally receive much lower Bayes factors. For the PT-simulated data set, the PT model is correctly discriminated from the CMF model (a fitted EPT model). We also see that although the ET model may not be correctly selected out when its BF is similar to that obtained for EP, PT, EPT, and CMF models, the ratios of the Bayes factors are close to unity. The much larger ratios between them for the real data validate our inference of the ET model (see section \ref{sec:comparison}).  Figure \ref{fig:BF_2D_sim} shows that the EP model is not favored over the Eccentricity model even when the former is the true model. However, the Eccentricity model is never favored on any of the real data sets, so this misidentification does not occur in practice. In conclusion, this discrimination test indicates that our identification (in section \ref{sec:LP}) of the PT model as the best model for the late Pleistocene is reliable.

We then apply the same test to the period 2-1\,Ma, which uses a different value of $\gamma$ as explained above. The results are shown in the right panel of Figure \ref{fig:BF_2D_sim}. We see that the correct model always has a larger Bayes factor than the other models. Yet we also see that for data simulated from the PT model, the CMF and EPT models have similar BFs as the PT model. However, as the PT model is not as fine tuned as the CMF model and has fewer adjustable parameters than the EPT model, we would generally invoke Occam's razor to select the PT model.

This experiment confirms that Bayesian model comparison and our interpretation of the Bayes factors allows us to select the correct model. We conclude that tilt (or obliquity) is the main ``pacemaker'' of the deglaciations over the last 2\,Myr, while precession may pace the deglaciations over the late Pleistocene. This indicates that precession becomes important in pacing terminations after the MPT. Other climate forcings, including GPI and inclination forcing, are unlikely to pace the deglaciations over the Pleistocene. 

\section{Sensitivity test}\label{sec:sensitivity}

We now perform a sensitivity test to check how sensitive a model's BF is to the choices of time scale and model priors. 

To do this we first change the time of the onset of the 100\,kyr cycles from 1\,Ma to 0.8\,Ma. We recalculate the BFs and show them in the lower left panel of Figure \ref{fig:BF_2D}. We see that the combination of obliquity and precession (i.e.\ the PT model) still paces the major terminations (DD, ML, and MS) better than obliquity alone. So our conclusion is robust to this change of the late-Pleistocene time span. 

\begin{figure*}[ht!]
  \centering
  \includegraphics[scale=0.5]{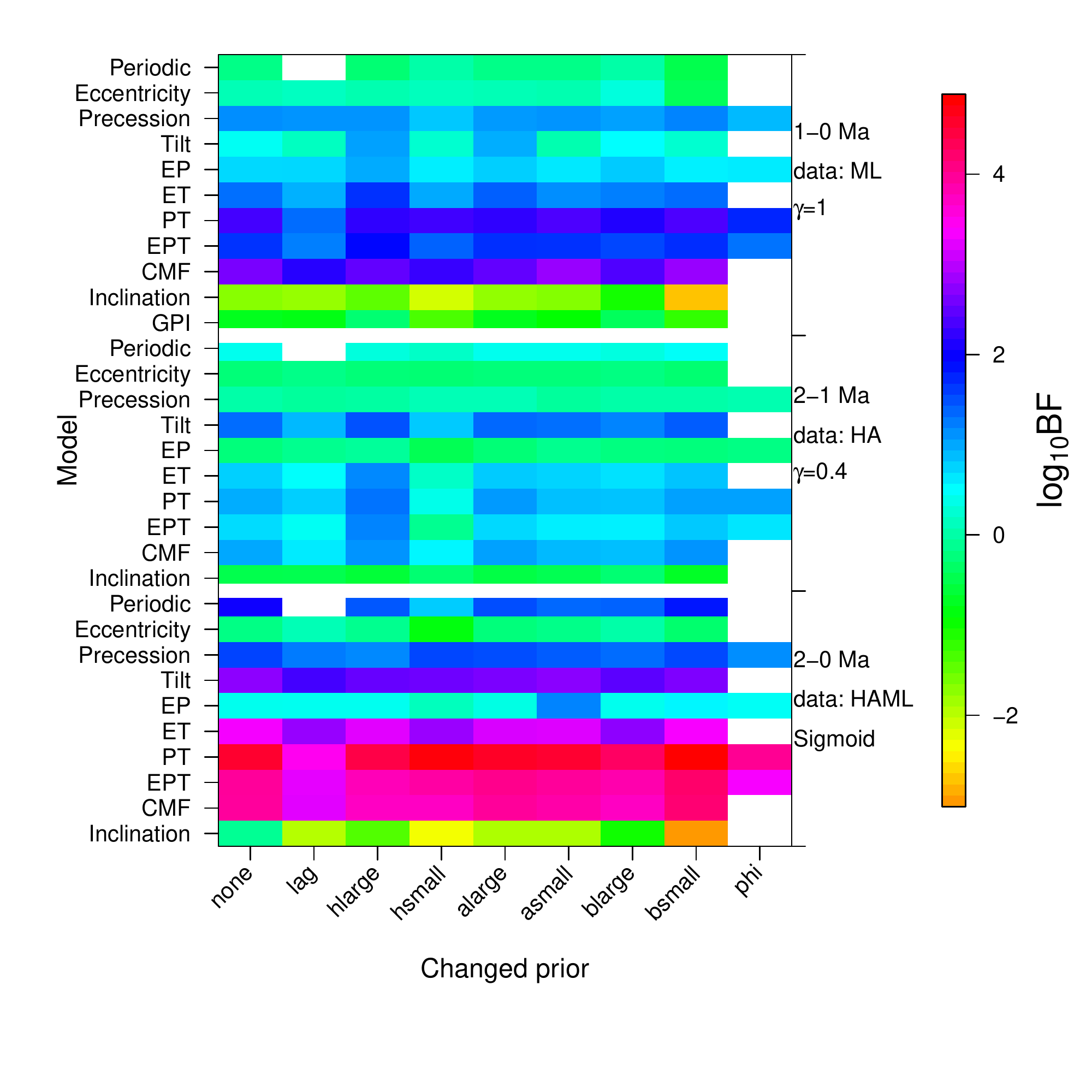}
  \vspace{-0.5in}
  \caption{Sensitivity test. Bayes factors for several models with a change in the range of priors (compared to what was used in section \ref{sec:BF} and Figure \ref{fig:BF_2D}). These are shown for three different data sets (and time scales) in the three blocks separated by white horizontal lines. In each block the logarithm of the Bayes factor is show on a color scale for each model (vertical axis) and change in prior (horizontal axis). The first column -- labeled `none' -- gives the BFs for models with the original priors for reference. Some models are not relevant for certain prior changes, so the corresponding slots are empty (white).  The three blocks are as follows. Top: pacing model with a constant background threshold with $\gamma=1$ for the ML data set (0--1\,Ma). Middle: pacing model with a constant background threshold with $\gamma=0.4$ for the HA data set (1--2\,Ma). Bottom: pacing model with a sigmoid background threshold for the HAML data set (0--2\,Ma).
}
  \label{fig:BF_2D_sen}
\end{figure*}

We then change the prior distributions over some model parameters and keep others fixed. We apply this sensitivity test to the ML, HA, and HAML data sets with time spans of 1--0\,Ma, 2--1\,Ma, and 2--0\,Ma, respectively. These three data sets are representative and conservative because they contain the major terminations as well as minor terminations identified in the HA data set, which is stacked from both benthic and planktic data sets. In each case we select the most favored types of the pacing model according to the model comparison in section \ref{sec:comparison}. They are: the constant background threshold with $\gamma=1$ for ML; the constant background threshold with $\gamma=0.4$ for HA; sigmoid background threshold for HAML. For each model, we change the range of the uniform prior on each parameter as follows (the name in parentheses is used to label the change in Figure \ref{fig:BF_2D_sen}) 

\begin{itemize} 
\item $\lambda=0 \rightarrow -10 \leq \lambda\leq 10$ ({\it lag}): Here we account for the possible time lag between the forcing and its effect (as was suggested by previous studies such as \citealt{hays76} and \citealt{imbrie80}).  $\lambda$ represents the time lag(s) of any model listed in Table \ref{tab:ts_models}, and $\lambda$ ranges -10 to 10\,kyr in steps of 1\,kyr. For models with a single component, a time lag is achieved by shifting the corresponding time series to the past or to the future. For compound models, each component is shifted independently, and the corresponding evidences are calculated by marginalizing the likelihood over time lags of all components.
\item $90\gamma<h_0<130\gamma \rightarrow 80\gamma<h_0<140\gamma \text{ ({\it hlarge}) and } 100\gamma<h_0<120\gamma$ ({\it hsmall}): We extend or shrink the upper and lower limits of the background threshold, $h_0$, by 10$\gamma$. Changing the prior distribution of $h_0$ is equivalent to changing the prior distribution of the period of a pacing model, because the average period is about $h_0+10-a$ (see section \ref{sec:pacing_constant}). The above changes only apply to models with $a\neq 0$ while the prior distribution of the Periodic model ($a=0$) is changed from $70\gamma<h_0<110\gamma$ first to $60\gamma<h_0<120\gamma$ and then to $80\gamma<h_0<100\gamma$. For models with a sigmoid trend, the prior distribution of $k$ is changed from $90<k<130$ first to $80<k<140$ and then to $100<k<120$. 
\item $15\gamma<a<35\gamma \rightarrow 5\gamma<a<45\gamma \text{ ({\it alarge}) or } 20\gamma<a<30\gamma$ ({\it asmall}): We extend or shrink the range of contribution factor of forcing, $a$, around its mean. These changes do not apply to the Periodic model, for which $a=0$.
\item $0<b<0.1 \rightarrow 0<b<0.2 \text{ ({\it blarge}) or } 0<b<0.05$ ({\it asmall}): We double or halve the upper limit of $b$, the contribution factor of the background in the termination model.
\item $\phi=0 \rightarrow -\pi<\phi<\pi$ ({\it phi}): We now allow any value for the the phase of the precession, $\phi$, which is related to the season of the insolation that forces the climate change.
\end{itemize}

The BFs for the models with each of the above changes are shown Figure \ref{fig:BF_2D_sen}, separated into three blocks corresponding to the different data sets, ML, HA, HAML. For the ML data set (1--0\,Ma; top block), the PT and CMF models are favored over the Tilt model for all changes in the priors. The PT and CMF models without time lags are also favored over corresponding models with lags. This indicates that the Tilt and Precession models pace climate change without significant time lags. Over the early Pleistocene (middle block), the Tilt model is marginally favored. The BFs of the EPT model vary a lot but are never higher than the Tilt model. For the HAML data set (2--0\,Myr; bottom block), the model combining a sigmoid trend and the PT forcing is favored for all changed priors. Moreover, the BF for the PT model increases when shrinking the range of the background fraction, $b$. The relative lack of significance of the background suggests a significant influence of obliquity and precession over the past 2\,Myr.

\begin{figure*}[ht!]
  \centering
  \includegraphics[scale=0.45]{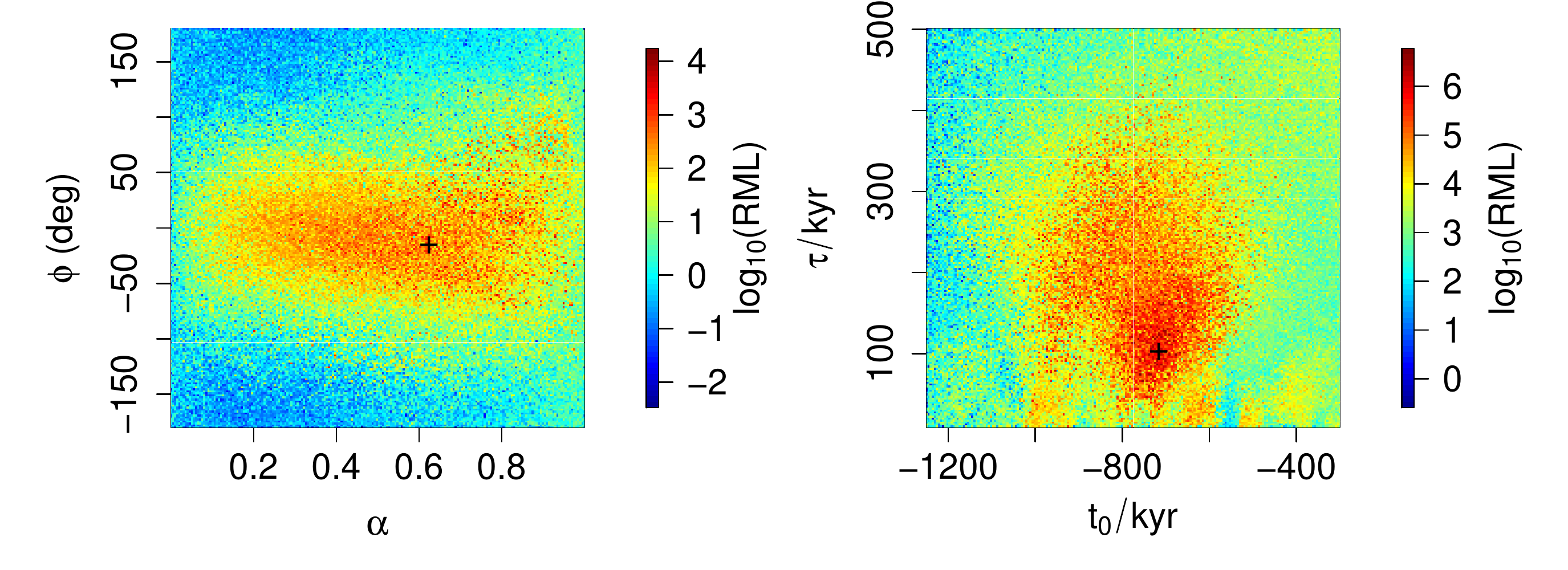}
  \caption{The distribution of the logarithm of the marginalized likelihood relative to the uniform model, $\log_{10}$(RML), for the PT model as a function of two model parameters. The left panel shows the distribution over the precession contribution factor ($\alpha$) and phase ($\phi$) for the PT model with $\gamma=1$ for the ML data set (the last 1\,Myr). The right panel shows the distribution over the transition time ($t_0$) and transition time scale ($\tau$) for the PT model with a sigmoid background threshold for the HAML data set (the last 2\,Myr). $10^6$ and $1.6\times 10^6$ sample points sampled in a $200\times 200$ grid were used to construct the left and right distributions respectively. For each panel, the most favored region is identified by applying a $25\times 25$ grid to the distribution, and is denoted by a cross.
Note that the scales saturate: likelihoods above or below the limits of the color bar are plotted using the extreme color.
  }
  \label{fig:likelihood_distribution}
\end{figure*}

To further investigate the role of precession in pacing the major late-Pleistocene deglaciations, we marginalize the likelihoods for the PT model over all its parameters except for the contribution factor of precession contribution factor, $\alpha$, and phase, $\phi$. (Note that the evidence is the likelihood marginalized over {\em all} model parameters.)  We do this for the ML data over the last 1\,Myr.  The distribution of this marginalized likelihood (relative to the uniform model) is shown in the left panel of Figure \ref{fig:likelihood_distribution}.  The highest values occur for phases ranging from $-50^\circ<\phi<+50^\circ$, indicating that the main pacemaker under this model is either the intensity of the northern hemisphere summer insolation or the duration of the southern summer (we cannot distinguish between these based on available data). While very small contribution factors, $\alpha<0.1$, are strongly disfavored, the model is otherwise not very sensitive to $\alpha$. Since $\alpha$ determines the size of the contribution of precession to the PT model (equation \ref{eqn:ts_function}), this means that some precession contribution is favored, but the exact amount is not well constrained. This broad high likelihood range of $\alpha$ and $\phi$ means that the pacing depends on the overall northern hemisphere summer insolation at a range of northern latitudes (or equivalently the duration of the southern summer) rather than that at a specific latitude and time in summer. This is consistent with \citealt{huybers11}'s conclusion that ``climate systems are thoroughly interconnected across temporal and spatial scales''.

We found in section \ref{sec:WP} that the pacing model with a sigmoid background threshold model was favored when modeling the whole Pleistocene. We now identify which parameters of that model are most favored by the data. To do this we calculate the marginalized likelihood (relative to the uniform model) for the PT model with a sigmoid background threshold as a function of both the transition time scale, $\tau$, and transition midpoint, $t_0$, on the HAML data set (i.e.\ we marginalize over all other parameters): see the right panel of Figure \ref{fig:likelihood_distribution}.
To explore this more completely we have extended the upper limit of $t_0$ from -700\,kyr to -300\,kyr. 
The peak is at around $\tau=100$\,kyr (about one glacial-interglacial cycle) and $t_0=-715$\,kyr. To visualize this transition, a sigmoid background model with this value of $\tau$ is shown in Figure \ref{fig:icevolume}. Defining the transition duration as the time taken for the ice volume to change from 25\% to 75\% of its maximum value, $\tau=100$\,kyr corresponds to a transition duration of 220\,kyr. This timescale for the MPT is consistent with the findings of \cite{honisch09,mudelsee97,tziperman03,martinez11}. It is shorter (more abrupt transition) that found by H07 and others \citep{raymo04,liu04,medina05,blunier98}, although Figure \ref{fig:likelihood_distribution} shows that longer time scales are not that improbable (but note that the likelihoods are shown on a logarithmic scale). The transition time of 715\,kyr ago is somewhat later than the mid-point of the MPT of $\sim$-900\,kyr identified by \cite{clark06} using a frequency spectrogram analysis. Yet our data/analysis permits a range of values, 
although we see that the region around -900\,kyr is disfavored for low values of $\tau$. Discrepancies from previous results could also arise from the fact that we use just termination data.

As a final sensitivity test, we change the sign of the contribution factor of forcing, $a$, to model possible anticorrelations between forcing models and the data over the late Pleistocene. 
We find that this significantly reduces the BF for all favored models, which shows that models with anticorrelations are a poor description of the data.

\section{Summary and conclusions}\label{sec:conclusion}

Using likelihood-based model comparison, we find that a combination of obliquity (axial tilt) and precession is the main pacemaker of the 12 major glacial terminations in the late Pleistocene. Obliquity alone can trigger minor terminations over the whole Pleistocene. The obliquity and precession pace the Pleistocene terminations without significant time lags, and their pacing roles can be identified with high significance.

We confirm the dominant role of obliquity in pacing the glacial terminations over the early Pleistocene. In contrast to
the conclusion of H07, we find that a model with obliquity alone describes the major and minor Pleistocene deglaciations (together) better than a model which combines obliquity with a trend in the background threshold.
Thus obliquity is sufficient to explain at least the time of minor terminations before and after the MPT, without reparameterizing the model as done by H07 and \cite{raymo97,paillard98,ashkenazy04,paillard04,clark06}. 

We observe that precession becomes important in pacing the $\sim$100\,kyr glacial-interglacial cycles after the MPT. Through the comparison of models with a linear trend and models with a sigmoid trend in the background threshold, we find that the glacial terminations over the whole Pleistocene can be paced by a combination of precession, obliquity, and a sigmoid trend in the background threshold. Using marginalized likelihoods, we find that the MPT has a time scale 
(the time required for ice volume to grow from 25\% to 75\% of the maximum)
of about 220\,kyr and a mid-point at around 715\,kyr before the present. This is rather late compared with other studies \citep{clark06}, although our data/analysis supports a broad range of values. Note that we do not assume the existence of a strict periodicity in the data, in contrast to some studies based on power spectrum analyses.
Since there is no significant change in the power spectrum of the insolation before and after the MPT, the MPT must be caused by a rapid change of {\em response} of the climate to the insolation, rather than by the insolation itself. This is consistent with previous studies \citep{paillard98,parrenin03,ashkenazy04,clark06}.

We also find that geomagnetic forcing and forcing by changes in the inclination of the Earth's orbital plane are unlikely to cause significant climate change over the last 2\,Myr. This weakens the 
suggestion that the Earth's orbital inclination relative to the invariable plane influences the climate \citep{muller97}. Our results also suggest that the modulation of cosmic rays or solar activity by the Earth's magnetic field has at best a limited impact on climate change on timescales between 10\,kyr and 1\,Myr, challenging the hypothesis that connects the geomagnetic paleointensity with climate change \citep{channell09}. 

The Bayesian modelling approach is well suited to multiple model comparison, because it evaluates all their evidences explicitly: a model is not selected just because some alternative ``noise'' model is rejected. Uncertainties in the data are also accommodated. Moreover, the approach automatically and consistently takes into account the model complexity, in contrast to most other methods (e.g.\ frequentist hypothesis testing, maximum likelihood ratio tests) which will favor more complex models unless they are penalized in some ad hoc way.

Our conclusions are reasonably robust to changes of parameters, priors, time scales, and data sets. The main uncertainty in our work comes from the identification of glacial terminations over the Pleistocene, although we have used different data sets of terminations to reduce this uncertainty. In future work, a more sophisticated Bayesian method (e.g.\ the method introduced by \citealt{bailer-jones12}) could be employed to model the full time series of climate proxies. Using this model inference approach, we may learn more about the mechanisms involved in the climate response to Milankovitch forcings. 

\section*{Acknowledgements}

We thank Joerg Lippold for pointing us to relevant literature, Marcus Christl for providing $^{10}$Be data, and Martin Frank for explaining the method of reconstructing the history of solar activity. Morgan Fouesneau, Eric Gaidos, and Gregor Seidel gave valuable comments on the manuscript. We also thank anonymous referee and the associate editor, Michel Crucifix, for their valuable comments. This work has been carried out as part of the Gaia Research for European Astronomy Training (GREAT-ITN) network. The research leading to these results has received funding from the European Union Seventh Framework Programme ([FP7/2007-2013] under grant agreement no.\ 264895.

\bibliographystyle{elsarticle-harv.bst}
\bibliography{climate}

\end{document}